\documentclass[11pt]{article}

\usepackage[final]{acl}

\usepackage{times}
\usepackage{latexsym}

\usepackage[T1]{fontenc}

\usepackage[utf8]{inputenc}

\usepackage{microtype}

\usepackage{inconsolata}

\usepackage{graphicx}

\usepackage{amsmath}
\usepackage{amssymb}
\usepackage{booktabs}
\usepackage{tabularx}
\usepackage{array}
\usepackage[table]{xcolor}
\usepackage{multirow}
\usepackage{siunitx}
\usepackage{dblfloatfix}
\usepackage{placeins} 
\usepackage{cuted}
\usepackage{float}
\title{Beyond the Individual: Virtualizing Multi-Disciplinary Reasoning for Clinical Intake via Collaborative Agents}

\author{
 \textbf{Huangwei Chen\textsuperscript{1,2,4}},
 \textbf{Wu Li\textsuperscript{1}},
 \textbf{Junhao Jia\textsuperscript{1,2,4}},
 \textbf{Yining Chen\textsuperscript{3}},
 \textbf{Xiaotao Pang\textsuperscript{2}},
 \textbf{Yalong Chen\textsuperscript{3}},
\\
 \textbf{Gonghui Li\textsuperscript{3}},
 \textbf{Haishuai Wang\textsuperscript{1,†}},
 \textbf{Jiajun Bu\textsuperscript{1}},
 \textbf{Lei Wu\textsuperscript{1,2†}},
\\
\\
 \textsuperscript{1}Zhejiang Key Laboratory of Accessible Perception and Intelligent Systems,\\
 College of Computer Science and Technology, Zhejiang University
\\
 \textsuperscript{2}Hangzhou Pujian Medical Technology Co., Ltd, China
\\
 \textsuperscript{3}Sir Run Run Shaw Hospital, Zhejiang University School of Medicine
\\
 \textsuperscript{4}School of Computer Science and Technology, Hangzhou Dianzi University
\\
\\
 \small{
   \textbf{Correspondence:} \href{haishuai.wang@zju.edu.cn}{haishuai.wang@zju.edu.cn}, \href{shenhai1895@zju.edu.cn}{shenhai1895@zju.edu.cn}
 }
}

\begin{document}
\maketitle
\begin{abstract}
The initial outpatient consultation is critical for clinical decision-making, yet it is often conducted by a single physician under time pressure, making it prone to cognitive biases and incomplete evidence capture. Although the Multi-Disciplinary Team (MDT) reduces these risks, they are costly and difficult to scale to real-time intake. We propose Aegle, a synchronous virtual MDT framework that brings MDT-level reasoning to outpatient consultations via a graph-based multi-agent architecture. Aegle formalizes the consultation state using a structured SOAP representation, separating evidence collection from diagnostic reasoning to improve traceability and bias control. An orchestrator dynamically activates specialist agents, which perform decoupled parallel reasoning and are subsequently integrated by an aggregator into a coherent clinical note. Experiments on ClinicalBench and a real-world RAPID-IPN dataset across 24 departments and 53 metrics show that Aegle consistently outperforms state-of-the-art proprietary and open-source models in documentation quality and consultation capability, while also improving final diagnosis accuracy. Our code is available at \url{https://github.com/HovChen/Aegle}.
\end{abstract}

\section{Introduction}
\label{sec:intro}
The trajectory of clinical care is fundamentally established during the initial consultation~\citep{starfield2005primarycare}. In this pivotal phase, a physician must transmute a patient's unstructured narrative of symptoms and concerns into a structured medical record, crystallizing it as the Initial Progress Note (IPN) in the SOAP (Subjective, Objective, Assessment, Plan) format. This document serves as more than a mere administrative summary; it is the cornerstone for downstream diagnostic decisions and treatment planning~\citep{krishna-etal-2021-generating}. Consequently, the comprehensiveness and accuracy of this intake process are paramount to effective healthcare delivery, serving as the bedrock upon which the entire clinical pathway rests.

\begin{figure}[t]
    \centering
    \includegraphics[width=\linewidth]{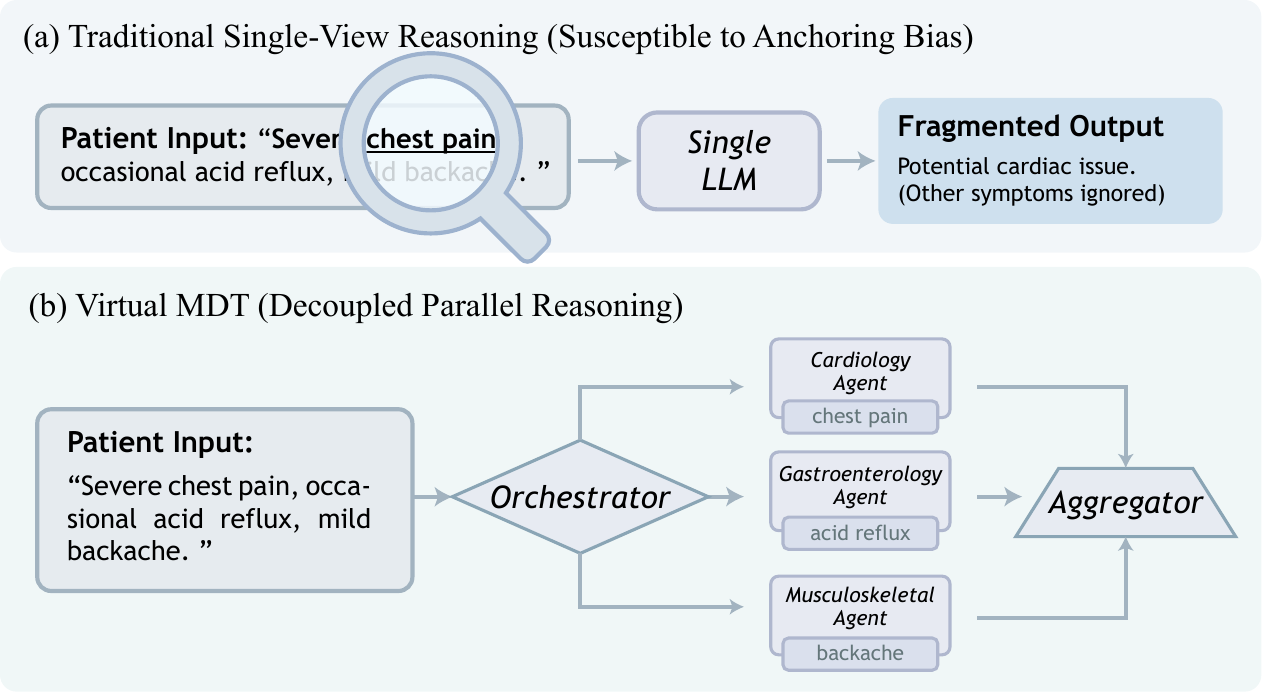}
    \caption{Single-view vs.\ virtual MDT reasoning for clinical intake. (a) A single LLM is prone to anchoring, over-focusing on salient symptoms, and producing fragmented notes. (b) A virtual MDT enables decoupled parallel specialist reasoning coordinated by an Orchestrator and integrated by an Aggregator, improving evidence coverage and coherence.}
    \label{fig:intro}
\end{figure}

However, achieving high-quality intake in routine practice is fraught with cognitive and systemic challenges~\citep{you2025ambient}. The traditional process typically relies on a single physician operating under significant time constraints. When formulating diagnoses while simultaneously engaging in empathetic dialogue, individual providers are susceptible to anchoring bias~\citep{croskerry2013mindful}, fixating on prominent symptoms while overlooking subtler, yet critical, diagnostic clues. As illustrated in Fig.~\ref{fig:intro}(a), this ``single-view'' setting narrows the exploration of the diagnostic space and can lead to fragmented evidence capture in the resulting note. This is not merely a matter of physician competence but a fundamental limit of human cognitive bandwidth when processing high-entropy patient narratives under time pressure.

To address complex cases where a single perspective is insufficient, medical practice traditionally turns to a Multi-Disciplinary Team (MDT)~\citep{taylor2010mdtcancer}. By aggregating specialists from diverse fields, MDT enables parallel and complementary reasoning across different clinical perspectives, mitigating the anchoring effects inherent to single-view decision making (Fig.~\ref{fig:intro}(b)). While this collaborative model significantly reduces the risk of oversight, it is inherently resource-intensive, asynchronous, and difficult to scale. Organizing a team of human experts for every routine outpatient consultation is logistically impractical. Thus, a critical gap remains: \textit{how can we transpose the systematic depth of MDT-level reasoning to the widespread, real-time outpatient intake phase without prohibitive resource costs?}

In this paper, we bridge this gap by proposing \textbf{Aegle}, a multi-agent framework that virtualizes the MDT paradigm. Rather than relying on a single Large Language Model (LLM) or a static chain of agents, Aegle introduces a novel computational architecture for medical inquiry. We posit that the essence of effective collaboration lies in \textit{decoupled parallel reasoning}, where distinct specialist agents analyze the case from their unique domain perspectives without interference, followed by a semantic aggregation phase. Furthermore, to address the efficiency issues common in multi-agent systems, we implement a dynamic topology controlled by a meta-cognitive orchestrator. This allows the system to adaptively scale its reasoning network based on the real-time completeness of the clinical documentation.

Our contributions are summarized as follows:

\begin{itemize}
    \item We propose \textbf{Aegle}, a Synchronous Virtual MDT framework that leverages decoupled parallel reasoning to transcend physical resource constraints. This paradigm transposes systematic, inpatient-level diagnostic depth into real-time outpatient inquiries, significantly enhancing robustness while mitigating single-view cognitive biases.

    \item We propose a \textbf{State-Aware Dynamic Topology} that aligns multi-agent collaboration with the evolving clinical document. By implementing on-demand specialist activation, this mechanism dynamically constructs reasoning paths tailored to case-specific ambiguity, thereby maximizing the diagnostic signal-to-noise ratio and ensuring high-density information gathering.

    \item We conduct a comprehensive evaluation across \textbf{24 clinical departments} using \textbf{53 fine-grained metrics}. Empirical results demonstrate Aegle's superiority over state-of-the-art baselines in diagnostic accuracy and documentation quality, establishing a robust benchmark for next-generation clinical AI assistants.
\end{itemize}

\section{Related Works}\label{sec:related_works}
\subsection{LLMs for Clinical Consultation and Documentation} 
LLMs have shown promise in clinical workflow optimization tasks such as clinical documentation support and conversational assistance during patient intake~\citep{zhou2025llmhealthcare}. In the realm of documentation, models function as semantic compressors, transforming unstructured dialogues into standardized formats like SOAP notes~\citep{krishna-etal-2021-generating}. While models such as Med-PaLM 2 have achieved accuracy comparable to human scribes in summarizing static records~\citep{singhal2025medpalm2}, they exhibit significant fragility in temporal reasoning. Specifically, when summarizing longitudinal patient trajectories, these models often succumb to the ``lost-in-the-middle'' effect, failing to accurately distinguish between historical ailments and current presenting symptoms, thereby compromising the integrity of the medical record~\citep{kruse-etal-2025-large, zeng2025trajcoa}.

Conversely, in interactive consultation, frameworks such as AMIE~\citep{tu2025amie} and Healthcare Agent~\citep{ren2025healthcareagent} have attempted to simulate the diagnostic inquiry process. Despite their conversational fluency, a critical limitation persists: these monolithic systems largely operate as passive information receivers~\citep{zhou2025activereasoning}. Rather than executing proactively asking rule-out questions to narrow the differential diagnosis space, they tend to hallucinate details or prematurely commit to a diagnosis based on incomplete user input~\citep{qiu2025diagagents}. This passivity reveals a fundamental misalignment with real-world intake, where the core challenge lies not merely in processing available text, but in the strategic elicitation of missing evidence~\citep{brooks2024stigmatizing}.

\subsection{Multi-Agent Systems for Clinical Reasoning}\label{subsec:mas_reasoning} 
To overcome the cognitive limits of single-model architectures, Multi-Agent Systems (MAS) have emerged as a promising paradigm for structured collaboration and distributed problem solving~\citep{hu2026landscape,shi2025realm,jia2026scimind,yu2025visual,yang2026from,xu2026rcbsf,zhang2025mmcnav,chen2025ccgseobench}. By assigning specialized roles such as oncologists, radiologists, and pathologists to distinct agents, frameworks like MedAgents~\citep{tang-etal-2024-medagents}, MAC~\citep{chen2025mac}, and MedCollab~\citep{zhan2026medcollab} leverage dialectical debate or role-specialized collaboration to decompose complex diagnostic tasks. Works like FetalAgents~\citep{hu2026fetalagents} and LungNoduleAgent~\citep{yang2026lungnoduleagent} demonstrate the potential of MAS in specialized clinical domains. RareAgents~\citep{chen2026rareagents} and DeepRare~\citep{zhao2026deeprare} extend MAS to rare-disease diagnosis and treatment support. MDAgents introduces an adaptive topology that dynamically structures collaboration based on the perceived medical complexity of the case, thereby optimizing the trade-off between accuracy and computational cost~\citep{kim2024mdagents}. Furthermore, frameworks such as ClinicalLab have demonstrated the utility of agentic collaboration in managing multi-departmental diagnostics, simulating the referral and consultation dynamics of a physical hospital~\citep{yan2025clinicallab}.

Related ideas have also been explored in the broader multi-agent literature. ChatEval applies multi-agent debate to evaluation rather than clinical reasoning~\citep{chan-etal-2024-chateval}, while sparse communication topologies have been shown to reduce redundant exchanges in debate-based systems~\citep{li-etal-2024-improving-multi}. DyLAN dynamically selects agent teams and interaction structures based on the task~\citep{liu-etal-2023-dylan}, and OSC studies cognitive orchestration through dynamic knowledge alignment in multi-agent collaboration~\citep{zhang-etal-2025-osc}. In contrast, Aegle targets interactive clinical intake, where coordination must remain grounded in an explicitly structured SOAP state and in a staged separation between evidence elicitation and diagnostic synthesis.

However, the ``black-box'' interaction between agents introduces varying degrees of collaborative failure modes. A recent large-scale audit of medical MAS reveals that agentic collaboration can lead to ``flawed consensus'' where agents reinforce each other's biases, and the suppression of correct minority opinions during the voting process~\citep{gu2025medagentaudit}. Additionally, the interplay of multiple probabilistic models creates a problem of ``compound opacity'', making it exponentially difficult to trace the provenance of a clinical error~\citep{salehi2025beyond}. These vulnerabilities highlight that while MAS can broaden the hypothesis space, they require rigid structural constraints to prevent unanchored speculation, a gap that our proposed framework specifically addresses.

\section{Methodology}
\label{sec:methodology}

\begin{figure*}
    \centering
    \includegraphics[width=1\linewidth]{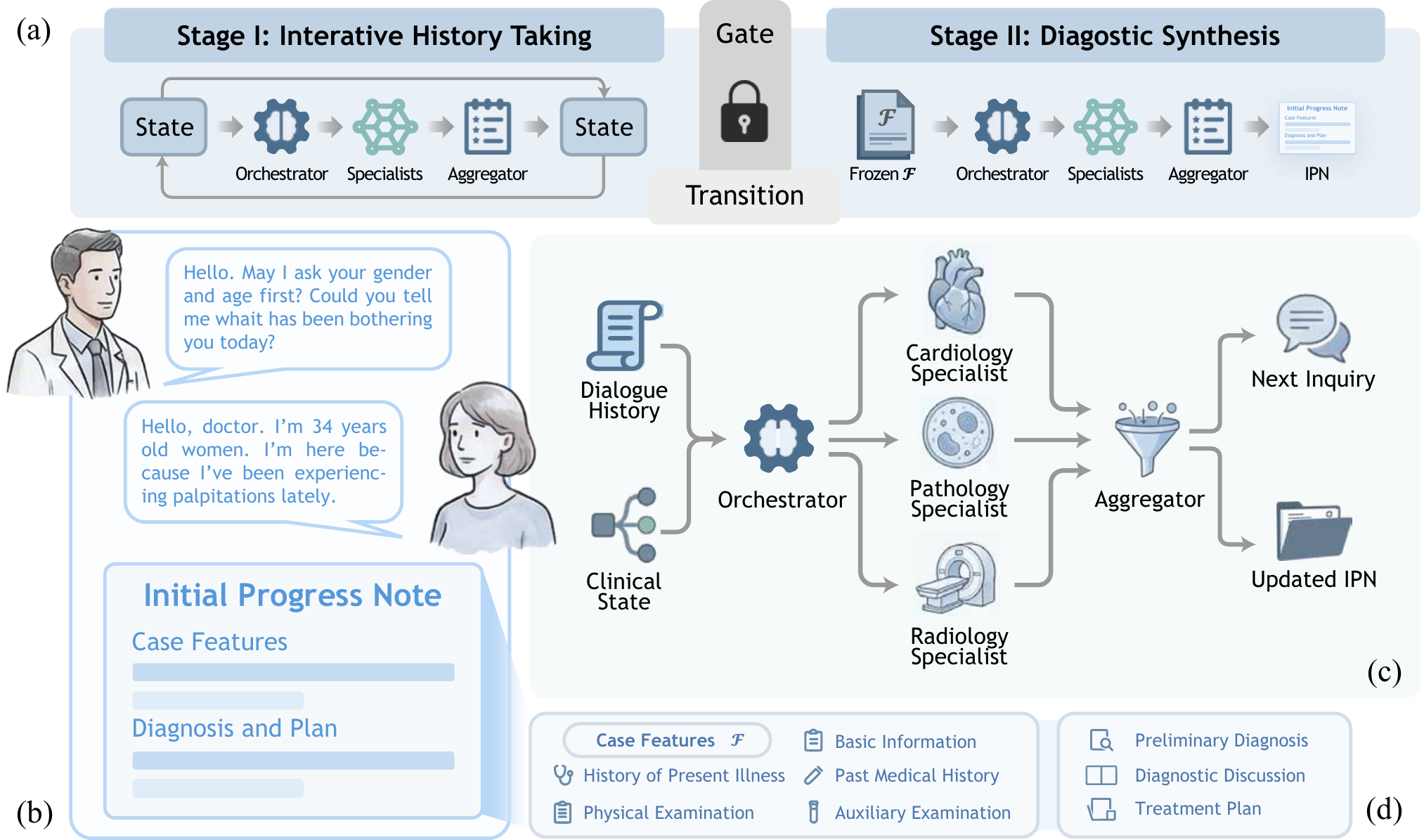}
    \caption{Overview of the Aegle framework.
    (a) A two-stage consultation workflow consisting of iterative history taking followed by diagnostic synthesis after freezing the case feature set $\mathcal{F}$.
    (b) An evolving draft Integrated Patient Note (IPN) that is incrementally updated throughout the consultation.
    (c) Dynamic multi-agent collaboration, where a context-aware Orchestrator activates relevant specialist agents and an Aggregator integrates their outputs to update the clinical state and generate the next inquiry.
    (d) Structured clinical state $\mathcal{S}_t = [\mathcal{F}_t, \mathcal{P}_t]$ separating evidentiary features from diagnostic and planning components.}
    \label{fig:framework}
\end{figure*}

We propose \textbf{Aegle}, a multi-agent consultation framework designed to virtualize the cognitive benefits of MDT collaboration during early-stage patient encounters. As illustrated in Fig.~\ref{fig:framework}, Aegle integrates structured, virtualized MDT-style collaboration directly into the consultation workflow. By coordinating multiple specialized agents during information gathering, the framework aims to surface overlooked considerations earlier and to reduce bias arising from single-perspective reasoning.

Built upon DeepSeek-V3.2~\citep{deepseekai2025deepseekv32}, Aegle instantiates a constrained graph-based agentic topology. Agent interactions are governed by explicit state representations and execution protocols, enabling controllable, transparent, and bias-aware clinical dialogue.

\subsection{Structured Clinical State}

To ground multi-agent collaboration in established clinical practice, Aegle formalizes the consultation state using the canonical SOAP schema. We denote the clinical state at turn $t$ as $\mathcal{S}_t$. Beyond its role as a documentation standard, SOAP provides a cognitive structure that explicitly separates evidence collection from diagnostic interpretation, thereby supporting bias-aware reasoning.

We decompose $\mathcal{S}_t$ into two functionally distinct components:
\begin{itemize}
    \item \textbf{Case Features ($\mathcal{F}$).}
    Corresponding to the \textit{Subjective} and \textit{Objective} sections of SOAP, $\mathcal{F}$ serves as an incremental repository of factual evidence. It accumulates verifiable patient information throughout the consultation, including Basic Information, History of Present Illness, Past Medical History, Physical Examination, and Auxiliary Examination results.

    \item \textbf{Diagnosis and Plan ($\mathcal{P}$).}
    Corresponding to the \textit{Assessment} and \textit{Plan} sections, $\mathcal{P}$ represents the analytical output of the consultation. It includes the preliminary diagnosis, diagnostic reasoning, and treatment plan, all of which are derived exclusively from the finalized case features in $\mathcal{F}$.
\end{itemize}

The structured state is defined as $\mathcal{S}_t = [\mathcal{F}_t, \mathcal{P}_t]$ and functions as a shared blackboard accessible to all agents. Aegle enforces a unidirectional dependency from $\mathcal{F}$ to $\mathcal{P}$ such that diagnostic and planning components may only be generated after evidence stabilization. This constraint explicitly links clinical conclusions to accumulated evidence, ensuring traceability and mitigating premature commitment to unsupported hypotheses.

\subsection{Multi-Agent Graph Topology}

Aegle operationalizes virtual MDT collaboration through a dynamic multi-agent graph topology composed of three types of nodes, each fulfilling a distinct role in the consultation workflow.

\paragraph{Orchestrator.}
The Orchestrator acts as a routing and coordination policy $\pi_{\text{orch}}$ that governs agent activation. It does not perform medical reasoning itself. Instead, it allocates computational attention by selecting a subset of specialist agents based on the evolving consultation context:
\begin{equation}
    A_{sub}, \iota = \pi_{\text{orch}}(\mathcal{H}_t, \mathcal{F}_t),
    \quad A_{sub} \subseteq \mathcal{A}_{\text{total}},
\end{equation}
where $\mathcal{H}_t$ denotes the dialogue history and $\iota$ specifies context-dependent task instructions. This selective activation mechanism mirrors real-world MDT practice by engaging specialized expertise only when warranted by the available evidence, thereby avoiding unnecessary or premature expert involvement during early-stage information gathering.

\paragraph{Specialist Agents.}
Each specialist agent operates as an independent domain expert, analyzing the clinical state from a distinct medical perspective. Specialists are executed in parallel and generate proposed updates to the clinical state in isolation. This decoupled architecture preserves hypothesis diversity by construction and delays consensus formation, reflecting the cognitive advantage of independent expert opinions in MDT discussions.

\paragraph{Aggregator.}
The Aggregator $\pi_{\text{agg}}$ serves as the interface between internal agent reasoning and patient-facing communication. It follows a write-then-speak protocol. First, it validates and integrates specialist proposals to update the structured clinical state:
\begin{equation}
    \mathcal{S}_{t+1} = \pi_{\text{agg}}^{\text{write}}\left(
    \mathcal{S}_t, \{\Delta \mathcal{S}_t^{(a)}\}_{a \in A_{sub}}
    \right).
\end{equation}
Subsequently, it generates the patient-facing utterance conditioned solely on the updated state:
\begin{equation}
    u_{t+1} = \pi_{\text{agg}}^{\text{speak}}(\mathcal{S}_{t+1}).
\end{equation}
This separation ensures internal consistency and technical precision of the medical record while maintaining clear and empathetic communication with the patient.

\subsection{Sequential Clinical Execution}

Building upon the structured clinical state and defined agent roles, Aegle executes consultations through a two-stage finite state machine. This temporal structure enforces a strict separation between evidence acquisition and diagnostic reasoning, serving as an explicit bias-control mechanism.

\begin{figure*}
    \centering
    \includegraphics[width=1\linewidth]{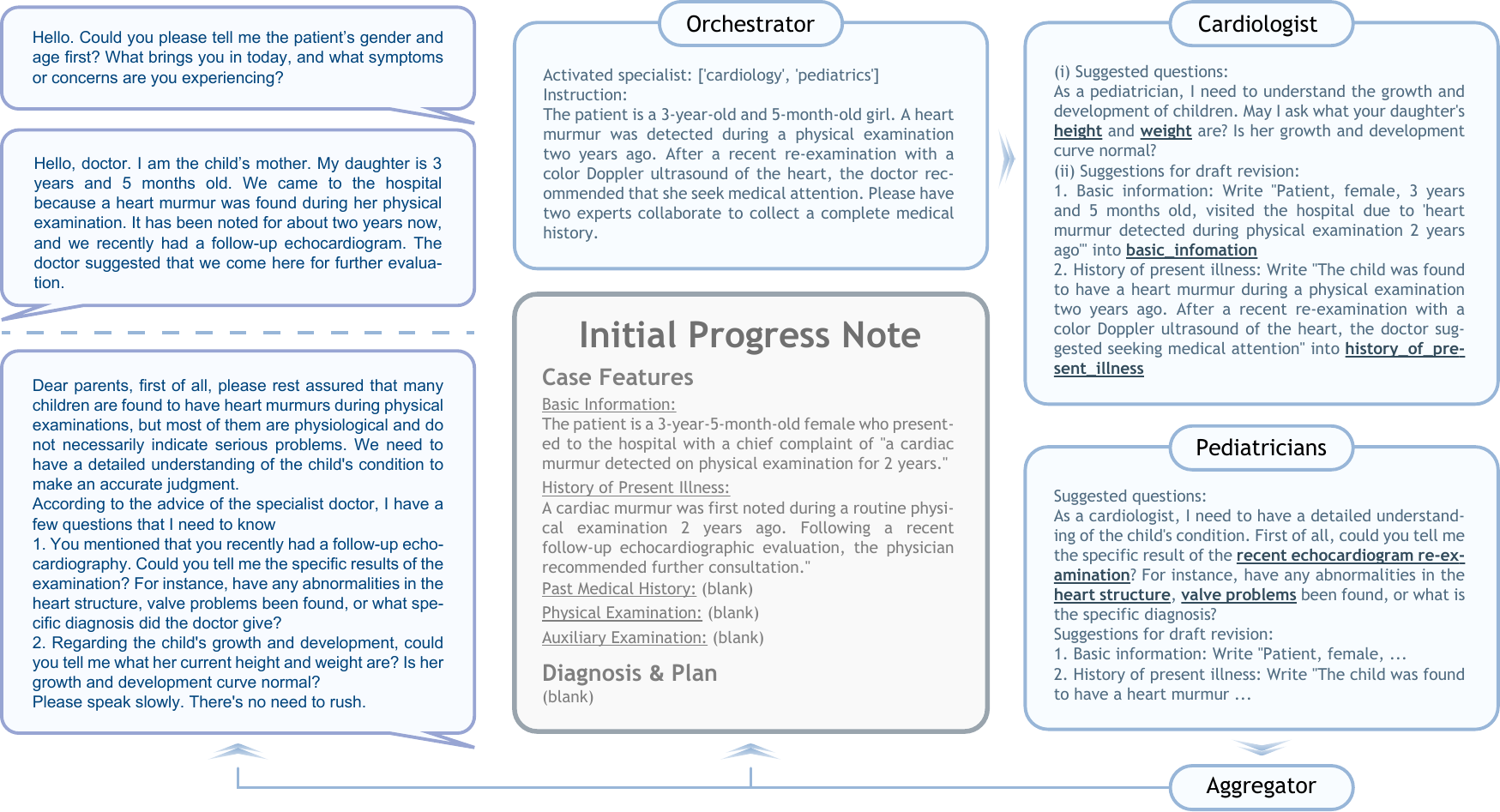}
    \caption{Stage I of Aegle, iterative history taking, illustrated with a pediatric heart murmur case. Patient responses are incorporated into the structured SOAP state and analyzed by multiple specialist agents in parallel. Each specialist proposes follow-up questions and targeted updates to case features. The Aggregator integrates these suggestions and generates the next patient-facing inquiry under a write-then-speak protocol.}
    \label{fig:historytaking}
\end{figure*}

\begin{table*}[h]
\centering

\resizebox{\textwidth}{!}{
\setlength{\tabcolsep}{3pt}
\begin{tabular}{
    l
    r c l  r c l  r c l  r c l
    c|c
    r c l  r c l  r c l  r c l
}
\toprule
\multirow{2}{*}{\textbf{Model}}
& \multicolumn{12}{c}{\textbf{ClinicalBench}}
& \multicolumn{2}{c}{}
& \multicolumn{12}{c}{\textbf{RAPID-IPN}} \\
\cmidrule(lr){2-13} \cmidrule(lr){16-27}
& \multicolumn{3}{c}{\textbf{IDEA}}
& \multicolumn{3}{c}{\textbf{SOAP}}
& \multicolumn{3}{c}{\textbf{READ}}
& \multicolumn{3}{c}{\textbf{chrF++}}
& &
& \multicolumn{3}{c}{\textbf{IDEA}}
& \multicolumn{3}{c}{\textbf{SOAP}}
& \multicolumn{3}{c}{\textbf{READ}}
& \multicolumn{3}{c}{\textbf{chrF++}} \\
\midrule

HuatuoGPT-o1-7B$^\dagger$$^\ddagger$
& 46.54 & $\pm$ & 11.16 & 37.98 & $\pm$ & 14.11 & 68.66 & $\pm$ & 6.30  & 9.62 & $\pm$ & 1.98
& &
& 41.89 & $\pm$ & 7.69  & 40.28 & $\pm$ & 12.96 & 68.58 & $\pm$ & 3.00  & 6.66 & $\pm$ & 1.23 \\

Qwen3-8B-think$^\dagger$
& 43.18 & $\pm$ & 9.22  & 27.66 & $\pm$ & 10.69 & 66.74 & $\pm$ & 5.48  & 10.19 & $\pm$ & 2.11
& &
& 39.93 & $\pm$ & 8.99  & 28.92 & $\pm$ & 13.42 & 66.14 & $\pm$ & 5.67  & 6.89 & $\pm$ & 1.73 \\

Baichuan-M2-32B$^\dagger$$^\ddagger$
& 38.02 & $\pm$ & 9.05  & 25.58 & $\pm$ & 10.63 & 64.00 & $\pm$ & 6.93  & 11.46 & $\pm$ & 2.45
& &
& 39.40 & $\pm$ & 9.89  & 31.54 & $\pm$ & 12.25 & 65.42 & $\pm$ & 5.81  & 9.24 & $\pm$ & 2.73 \\

Lingshu-32B$^\dagger$$^\ddagger$
& 50.50 & $\pm$ & 10.62 & 39.76 & $\pm$ & 14.92 & 71.11 & $\pm$ & 5.98  & 11.53 & $\pm$ & 2.44
& &
& 51.41 & $\pm$ & 10.66 & 47.89 & $\pm$ & 13.56 & 72.07 & $\pm$ & 3.94  & 8.62 & $\pm$ & 1.89 \\

DeepSeek-V3.2$^\dagger$
& 50.51 & $\pm$ & 8.61  & 38.64 & $\pm$ & 12.57 & 71.73 & $\pm$ & 4.62  & 17.33 & $\pm$ & 2.62
& &
& 54.35 & $\pm$ & 8.76  & 47.39 & $\pm$ & 10.95 & 72.14 & $\pm$ & 4.56  & 14.09 & $\pm$ & 2.69 \\

DeepSeek-V3.2-Thinking$^\dagger$
& 46.73 & $\pm$ & 9.68  & 34.62 & $\pm$ & 12.71 & 69.53 & $\pm$ & 4.90  & 16.17 & $\pm$ & 2.80
& &
& 49.37 & $\pm$ & 11.12 & 41.98 & $\pm$ & 13.42 & 70.72 & $\pm$ & 4.69  & 12.56 & $\pm$ & 2.92 \\

GLM-4.6$^\dagger$
& 47.02 & $\pm$ & 8.96  & 34.70 & $\pm$ & 11.87 & 68.48 & $\pm$ & 5.00  & 16.73 & $\pm$ & 2.43
& &
& 48.23 & $\pm$ & 11.47 & 40.44 & $\pm$ & 13.68 & 68.92 & $\pm$ & 5.43  & 13.31 & $\pm$ & 2.82 \\

Kimi~K2~Thinking$^\dagger$
& 54.05 & $\pm$ & 9.57  & 42.86 & $\pm$ & 12.70 & 70.76 & $\pm$ & 6.51  & 17.10 & $\pm$ & 2.89
& &
& 55.26 & $\pm$ & 10.00 & 49.45 & $\pm$ & 10.79 & 70.55 & $\pm$ & 5.77  & 13.92 & $\pm$ & 2.60 \\

MiniMax-M2$^\dagger$
& 57.78 & $\pm$ & 11.02 & 46.18 & $\pm$ & 12.46 & 73.87 & $\pm$ & 7.05  & 16.41 & $\pm$ & 2.90
& &
& 63.01 & $\pm$ & 10.91 & 56.84 & $\pm$ & 10.52 & 79.74 & $\pm$ & 9.28  & 14.81 & $\pm$ & 3.43 \\

GPT-4o$^\star$
& 41.05 & $\pm$ & 9.73  & 29.38 & $\pm$ & 12.75 & 67.66 & $\pm$ & 5.20  & 10.83 & $\pm$ & 1.98
& &
& 44.70 & $\pm$ & 9.80  & 34.79 & $\pm$ & 13.51 & 69.89 & $\pm$ & 3.66  & 10.84 & $\pm$ & 2.14 \\

Gemini~2.5$^\star$
& 48.35 & $\pm$ & 11.45 & 35.58 & $\pm$ & 15.63 & 70.10 & $\pm$ & 9.38  & 17.82 & $\pm$ & 3.34
& &
& 49.89 & $\pm$ & 11.10 & 39.03 & $\pm$ & 15.49 & 71.25 & $\pm$ & 4.75  & 14.69 & $\pm$ & 3.14 \\

Qwen3-Max$^\star$
& 61.75 & $\pm$ & 9.56  & 53.40 & $\pm$ & 11.25 & 74.99 & $\pm$ & 4.57  & 17.73 & $\pm$ & 2.48
& &
& 60.82 & $\pm$ & 7.83  & 57.84 & $\pm$ & 8.53  & 74.29 & $\pm$ & 4.19  & 14.66 & $\pm$ & 2.42 \\

Doubao-Seed-1.6$^\star$
& 51.51 & $\pm$ & 8.85  & 39.80 & $\pm$ & 11.20 & 69.03 & $\pm$ & 5.30  & 18.04 & $\pm$ & 2.72
& &
& 51.11 & $\pm$ & 8.90  & 44.37 & $\pm$ & 9.95  & 68.35 & $\pm$ & 4.63  & 14.38 & $\pm$ & 2.61 \\

ERNIE-5.0-Preview$^\star$
& 47.09 & $\pm$ & 9.99  & 35.55 & $\pm$ & 11.93 & 67.19 & $\pm$ & 5.73  & 15.82 & $\pm$ & 2.69
& &
& 45.05 & $\pm$ & 9.49  & 40.03 & $\pm$ & 11.59 & 67.03 & $\pm$ & 5.13  & 12.41 & $\pm$ & 2.67 \\

\rowcolor[HTML]{F4F7F9}
CoT
& 64.72 & $\pm$ & 9.01
& 60.34 & $\pm$ & 7.09
& 76.94 & $\pm$ & 4.74
& \underline{25.83} & $\pm$ & 2.97
& &
& 65.61 & $\pm$ & 7.57
& 61.89 & $\pm$ & 7.54
& 78.57 & $\pm$ & 4.95
& 21.23 & $\pm$ & 3.55 \\

\rowcolor[HTML]{F4F7F9}
ToT
& 66.53 & $\pm$ & 9.75
& \underline{62.34} & $\pm$ & 6.64
& \underline{78.46} & $\pm$ & 5.49
& \textbf{25.96} & $\pm$ & 3.14
& &
& \underline{67.58} & $\pm$ & 6.85
& \textbf{63.96} & $\pm$ & 6.17
& 79.86 & $\pm$ & 5.50
& 21.77 & $\pm$ & 3.13 \\

\rowcolor[HTML]{EDF3F8}
MDAgents
& \underline{68.41} & $\pm$ & 9.61
& 56.45 & $\pm$ & 8.53
& \textbf{78.81} & $\pm$ & 6.76
& 24.83 & $\pm$ & 2.53
& &
& 66.58 & $\pm$ & 10.05
& 60.98 & $\pm$ & 8.92
& \textbf{79.95} & $\pm$ & 8.41
& 21.28 & $\pm$ & 2.73 \\

\rowcolor[HTML]{EDF3F8}
MedAgents
& 59.85 & $\pm$ & 9.39
& 54.48 & $\pm$ & 7.95
& 72.00 & $\pm$ & 5.88
& 23.66 & $\pm$ & 2.33
& &
& 61.51 & $\pm$ & 8.81
& 57.19 & $\pm$ & 7.42
& 74.30 & $\pm$ & 5.17
& \underline{22.04} & $\pm$ & 2.33 \\

\rowcolor[HTML]{EDF3F8}
\textbf{Aegle (Ours)}
& \textbf{72.78} & $\pm$ & 10.16
& \textbf{63.02} & $\pm$ & 5.57
& 77.55 & $\pm$ & 5.41
& \underline{25.83} & $\pm$ & 2.55
& &
& \textbf{71.52} & $\pm$ & 8.35
& \underline{63.92} & $\pm$ & 4.73
& \underline{79.93} & $\pm$ & 6.90
& \textbf{24.24} & $\pm$ & 2.44 \\

\bottomrule
\end{tabular}
}
\caption{Documentation quality evaluation on ClinicalBench and RAPID-IPN. Unshaded rows report frontier model comparisons across proprietary and open-source baselines. Shaded rows denote the comparison setting with matched base models, where light gray rows denote reasoning-strategy baselines and light blue rows denote medical multi-agent systems. Models marked with $\dagger$ are open-source, those with $\ddagger$ are medical-domain models, and those with $\star$ are proprietary models. Mean $\pm$ standard deviation is reported.}
\label{tab:doc_quality}
\end{table*}

\paragraph{Stage I: Iterative History Taking.}
As shown in Fig.~\ref{fig:historytaking}, the consultation begins with iterative history taking. The Orchestrator activates relevant specialist agents based on patient responses and the current completeness of $\mathcal{F}_t$. Each specialist examines the updated clinical state from its domain perspective and proposes follow-up questions together with evidence-centric revisions to the draft Integrated Patient Note.

The Aggregator integrates these parallel proposals, updates the case features in $\mathcal{F}$, and generates the next consultation question. This process continues until all mandatory fields in $\mathcal{F}$ are either populated or explicitly marked as unavailable by the patient, ensuring that downstream diagnostic reasoning is grounded in sufficient evidence.

\paragraph{Stage II: Diagnostic Synthesis.}
Once $\mathcal{F}$ is finalized, it is frozen to prevent further modification, and the system transitions deterministically to diagnostic synthesis. In this stage, the Orchestrator commissions specialist agents to perform independent diagnostic reasoning based on the same fixed evidentiary substrate. Specialists propose diagnostic hypotheses and treatment considerations without introducing new inquiries.

The Aggregator then integrates these heterogeneous perspectives to produce the final diagnosis and plan $\mathcal{P}$, resolving inconsistencies and generating a complete and coherent SOAP note. This staged execution ensures that diagnostic conclusions are derived exclusively from stabilized evidence, reinforcing traceability and reducing bias induced by early hypothesis fixation.

\section{Experiments}
\label{sec:setup}

\subsection{Datasets}
\label{subsec:datasets}
To evaluate Aegle's performance across diverse clinical scenarios, we utilize two distinct datasets:

\paragraph{ClinicalBench.} 
We employ ClinicalBench~\citep{yan2025clinicallab}, a comprehensive end-to-end benchmark derived from de-identified electronic health records (EHRs) of top-tier Grade 3A hospitals in China. It contains 1,500 cases covering 24 clinical departments and 150 diseases. Crucially, it enforces a strict data-leakage-free protocol and supports open-ended generation tasks, simulating the whole trajectory from triage to treatment planning.

\paragraph{RAPID-IPN.} 
To evaluate complex differential diagnosis, we curated the \textbf{R}eal-world \textbf{A}bdominal \textbf{P}ain \textbf{I}ntegrated \textbf{D}iagnostic-pathway \textbf{I}nitial \textbf{P}rogress \textbf{N}ote (RAPID-IPN) dataset from a top-tier Grade 3A hospital in Eastern China. Spanning from 2018 to 2024, this cohort comprises 322 patients with abdominal pain across 12 departments, encompassing internal medicine (e.g., Cardiology, Gastroenterology) and surgery (e.g., Hepatobiliary Surgery). A rigorous review protocol by three senior physicians ($>$5 years of experience) ensured that the standardized SOAP notes and treatment plans strictly aligned with patients' actual clinical trajectories, thereby guaranteeing real-world fidelity. Due to patient privacy regulations and institutional data governance policies, the RAPID-IPN dataset cannot be publicly released. All data were fully de-identified in accordance with local regulations prior to use, and the study protocol was reviewed and approved by the hospital’s ethics committee. 

\subsection{Experimental Setup}
\label{subsec:metrics}

\paragraph{Baselines and Evaluation Paradigm.} To rigorously benchmark our framework, we report two complementary comparison settings. We first compare Aegle against a broad set of proprietary and open-source single-LLM baselines to situate its performance among frontier models. We then conduct a fixed-backbone comparison in which CoT~\citep{wei2022cot}, ToT~\citep{yao2023tot}, MDAgents~\citep{kim2024mdagents}, MedAgents~\citep{tang-etal-2024-medagents}, and Aegle all use DeepSeek-V3.2, allowing us to isolate the effect of reasoning and collaboration structure. All rubric-based results are evaluated under an LLM-as-a-judge paradigm using gpt-4o-mini, with identical scoring prompts across all conditions to ensure fair and consistent comparison. We further conducted a small-scale human evaluation to validate the reliability of the LLM-as-a-judge paradigm; details are provided in Appendix~\ref{app:llm_judge_corr}.

\paragraph{Evaluation Metrics.}
We adopt a multi-dimensional evaluation framework that assesses both the consultation process and the resulting clinical documentation. Specifically, documentation quality is evaluated along clinical reasoning (IDEA), documentation standardization (SOAP), readability (READ), and surface-level similarity (chrF++). Consultation capability is assessed using a consultation skills rubric covering inquiry skills and humanistic care. To complement these rubric-based assessments with an objective correctness signal, we additionally report final diagnosis accuracy on ClinicalBench. Detailed metric introductions and rubrics are provided in Appendix~\ref{app:metrics}.

\subsection{Documentation Quality Evaluation}
\label{subsec:doc_quality}

\begin{table*}[ht]
\centering
\resizebox{\textwidth}{!}{%
\begin{tabular}{lcccccccc|cccccccc}
\toprule
\multirow{2}{*}{\textbf{Model}} 
& \multicolumn{7}{c}{\textbf{ClinicalBench}} 
& \multicolumn{2}{c}{}
& \multicolumn{7}{c}{\textbf{RAPID-IPN}} \\
\cmidrule(lr){2-8} \cmidrule(lr){11-17}
& \textbf{CA} & \textbf{QT} & \textbf{VER} & \textbf{PJ} & \textbf{SP} & \textbf{AB} & \textbf{Turns}
&&& \textbf{CA} & \textbf{QT} & \textbf{VER} & \textbf{PJ} & \textbf{SP} & \textbf{AB} & \textbf{Turns} \\
\midrule

HuatuoGPT-o1-7B$^\dagger$$^\ddagger$
& 4.01 & 4.44 & 4.88 & 4.15 & 4.96 & 4.92 & 21.67
&&& 4.00 & 4.43 & 4.89 & 4.20 & 4.99 & 4.98 & 19.42 \\

Qwen3-8B-think$^\dagger$
& 4.00 & 4.10 & 4.69 & 4.78 & 4.94 & 4.14 & 7.03
&&& 4.00 & 4.07 & 4.66 & 4.68 & 4.96 & 4.11 & 6.86 \\

Baichuan-M2-32B$^\dagger$$^\ddagger$
& 3.99 & 3.97 & 4.47 & 4.86 & 4.89 & 3.97 & 11.54
&&& 4.00 & 3.87 & 4.32 & 4.91 & 4.86 & 3.94 & 11.10 \\

Lingshu-32B$^\dagger$$^\ddagger$
& 4.00 & 4.52 & 4.91 & 4.34 & 4.98 & 4.91 & 8.03
&&& 4.00 & 4.48 & 4.88 & 4.30 & 4.99 & 4.92 & 7.36 \\

DeepSeek-V3.2$^\dagger$
& 4.01 & 4.55 & 4.81 & 4.31 & 4.93 & 4.86 & 20.50
&&& 4.02 & 4.64 & 4.79 & 4.25 & 4.95 & 4.90 & 19.61 \\

DeepSeek-V3.2-Thinking$^\dagger$
& 4.00 & 4.86 & 4.96 & 4.57 & 4.99 & 4.97 & 8.80
&&& 4.00 & 4.84 & 4.92 & 4.40 & 5.00 & 4.98 & 8.26 \\

GLM-4.6$^\dagger$
& 4.00 & 4.65 & 4.97 & 4.79 & 5.00 & 4.98 & 8.21
&&& 4.00 & 4.60 & 4.94 & 4.64 & 5.00 & 4.98 & 8.10 \\

Kimi K2 Thinking$^\dagger$
& 4.02 & 4.88 & 4.93 & 4.58 & 4.99 & 4.97 & 8.96
&&& 4.01 & 4.85 & 4.87 & 4.38 & 5.00 & 4.97 & 8.14 \\

MiniMax-M2$^\dagger$
& 4.04 & 4.77 & 4.77 & 4.44 & 4.99 & 4.85 & 9.14
&&& 4.04 & 4.34 & 4.05 & 4.68 & 5.00 & 4.57 & 21.23 \\

GPT-4o$^\star$
& 4.00 & 4.74 & 4.93 & 4.91 & 5.00 & 4.56 & 4.60
&&& 4.00 & 4.88 & 4.94 & 4.84 & 5.00 & 4.49 & 3.84 \\

Gemini 2.5$^\star$
& 3.88 & 3.58 & 4.12 & 4.89 & 4.38 & 4.00 & 29.98
&&& 3.95 & 3.60 & 4.09 & 4.89 & 4.28 & 3.96 & 30.00 \\

Qwen3-Max$^\star$
& 4.01 & 4.90 & 4.96 & 4.38 & 5.00 & 4.96 & 6.42
&&& 4.01 & 4.84 & 4.97 & 4.34 & 5.00 & 4.96 & 5.41 \\

Doubao-Seed-1.6$^\star$
& 4.00 & 4.45 & 4.88 & 4.61 & 4.99 & 4.66 & 10.43
&&& 4.00 & 4.32 & 4.82 & 4.56 & 5.00 & 4.66 & 10.04 \\

ERNIE-5.0-Preview$^\star$
& 4.00 & 4.55 & 4.92 & 4.59 & 4.97 & 4.39 & 5.06
&&& 4.00 & 4.36 & 4.84 & 4.47 & 4.97 & 4.36 & 4.51 \\

\rowcolor[HTML]{F4F7F9}
CoT
& 4.00 & 4.83 & 4.63 & 4.05 & 4.99 & 4.85 & 6.40
&&& 4.00 & 4.78 & 4.48 & 4.28 & 5.00 & 4.86 & 6.66 \\

\rowcolor[HTML]{F4F7F9}
ToT
& 4.00 & 4.86 & 4.65 & 4.00 & 4.99 & 4.89 & 7.37
&&& 4.00 & 4.84 & 4.56 & 4.15 & 5.00 & 4.91 & 7.91 \\

\rowcolor[HTML]{EDF3F8}
MDAgents
& 4.00 & 4.66 & 4.85 & 4.32 & 4.98 & 4.93 & 6.59
&&& 4.00 & 4.76 & 4.94 & 4.57 & 5.00 & 4.71 & 6.61 \\

\rowcolor[HTML]{EDF3F8}
MedAgents
& 4.00 & 4.58 & 4.90 & 4.55 & 5.00 & 4.80 & 6.96
&&& 4.00 & 4.62 & 4.92 & 4.57 & 5.00 & 4.71 & 6.61 \\

\rowcolor[HTML]{EDF3F8}
\textbf{Aegle (Ours)} 
& 4.02 & 4.95 & 4.94 & 4.21 & 5.00 & 5.00 & 10.16
&&& 4.03 & 4.96 & 4.93 & 4.19 & 5.00 & 5.00 & 8.84 \\
\bottomrule
\end{tabular}
}
\caption{Consultation Capability Evaluation Results on ClinicalBench and RAPID-IPN. Unshaded rows report frontier model comparisons across proprietary and open-source baselines. Shaded rows denote the comparison setting with matched base models, where light gray rows denote reasoning-strategy baselines and light blue rows denote medical multi-agent systems. Metrics: CA = Conversation Arrangement; QT = Question Types; VER = Verifications; PJ = Professional Jargon; SP = Speech; AB = Amiable Behavior; Turns = number of dialogue turns. Models marked with $\dagger$ are open-source, those with $\ddagger$ are medical-domain models, and those with $\star$ are proprietary models.}
\label{tab:cons_capability}
\end{table*}

The documentation quality results in Table~\ref{tab:doc_quality} reveal a clear and consistent performance advantage for Aegle across both evaluation settings. In the frontier model comparison, Aegle remains competitive with strong proprietary and open-source models. In the fixed-backbone comparison, it also outperforms strong reasoning-strategy baselines (CoT and ToT) as well as existing medical multi-agent systems (MDAgents and MedAgents). The gains are most pronounced in metrics related to internal coherence and evidential grounding, reflecting a shift from surface-level summarization toward more structured clinical reasoning.

In single-model baselines, IPN often exhibit a familiar failure mode: fluent narratives that read well locally but lack global alignment between history, assessment, and plan. While CoT-style prompting partially alleviates this issue by improving local reasoning consistency, it does not explicitly constrain how evidence is accumulated and reused across sections. Aegle mitigates this limitation by enforcing an explicit separation between case feature accumulation and diagnostic synthesis. As a result, diagnostic conclusions and management plans are consistently traceable to previously documented evidence, reducing omissions.

On metrics such as READ and chrF++, Aegle is comparable to the strongest reasoning baselines on both datasets, suggesting that linguistic fluency has largely saturated among modern LLMs. Consequently, further gains in clinical documentation quality depend less on wording and more on how information is structured, prioritized, and constrained.

\subsection{Diagnostic Accuracy}
\label{subsec:diag_accuracy}

To complement the rubric-based note-quality metrics with an objective correctness signal, we additionally evaluate final diagnosis accuracy on ClinicalBench, where standardized diagnostic labels are available. We compare Aegle against DeepSeek-V3.2, CoT, ToT, MedAgents, and MDAgents under a shared DeepSeek-V3.2 backbone. Unlike the original ClinicalBench formulation, our evaluation starts from the consultation phase: the model must first elicit evidence through interaction before producing a diagnosis, rather than being given the complete post-consultation record as input. We therefore report these results as end-to-end diagnosis accuracy under the clinical intake setting.

\begin{table}[t]
\centering
\small
\begin{tabular}{lc}
\toprule
\textbf{Method} & \textbf{Acc. (\%)} \\
\midrule
DeepSeek-V3.2 & 25.60 \\
CoT & 39.60 \\
ToT & 38.00 \\
MDAgents & 25.73 \\
MedAgents & 39.20 \\
\textbf{Aegle (Ours)} & \textbf{46.93} \\
\bottomrule
\end{tabular}
\caption{Final diagnosis accuracy on ClinicalBench. All compared methods use DeepSeek-V3.2.}
\label{tab:diag_accuracy}
\end{table}

As shown in Table~\ref{tab:diag_accuracy}, Aegle achieves the highest diagnosis accuracy, outperforming the underlying DeepSeek-V3.2 model by 21.33 points and surpassing both reasoning-strategy baselines and prior medical multi-agent systems. This result is consistent with the IDEA and SOAP gains in Table~\ref{tab:doc_quality}, indicating that Aegle improves not only note structure but also final diagnostic decisions under the same backbone.

\subsection{Consultation Capability Evaluation}
\label{subsec:consultation_eval}

The consultation capability results in Table~\ref{tab:cons_capability} show that Aegle’s strengths are mainly reflected in how consultation information is elicited and validated, rather than in stylistic or expressive aspects of dialogue. Across both benchmarks, Aegle consistently demonstrates a more directed questioning strategy, in which dialogue turns are organized around resolving clinically relevant uncertainties.

When compared with single-model baselines and reasoning-strategy baselines such as CoT and ToT, Aegle exhibits a more structured pattern of information verification. Its higher VER and QT scores indicate that follow-up questions are more frequently used to confirm or refine patient-provided information, instead of extending the conversation through loosely related prompts. This pattern is characteristic of specialist-driven history taking, where each question serves a specific diagnostic purpose.

Conversational style and humanistic expression metrics remain close to their upper bounds for most competitive models, which limits their discriminative value. In terms of dialogue length, Aegle occupies a middle range, avoiding both very short interactions that may under-verify critical details and excessively long exchanges that dilute diagnostic focus. By maintaining this balance while achieving strong verification and question coverage, Aegle demonstrates a consultation behavior that is both efficient and clinically grounded, which is consistent across the two benchmarks.

\begin{figure*}
    \centering
    \includegraphics[width=1\linewidth]{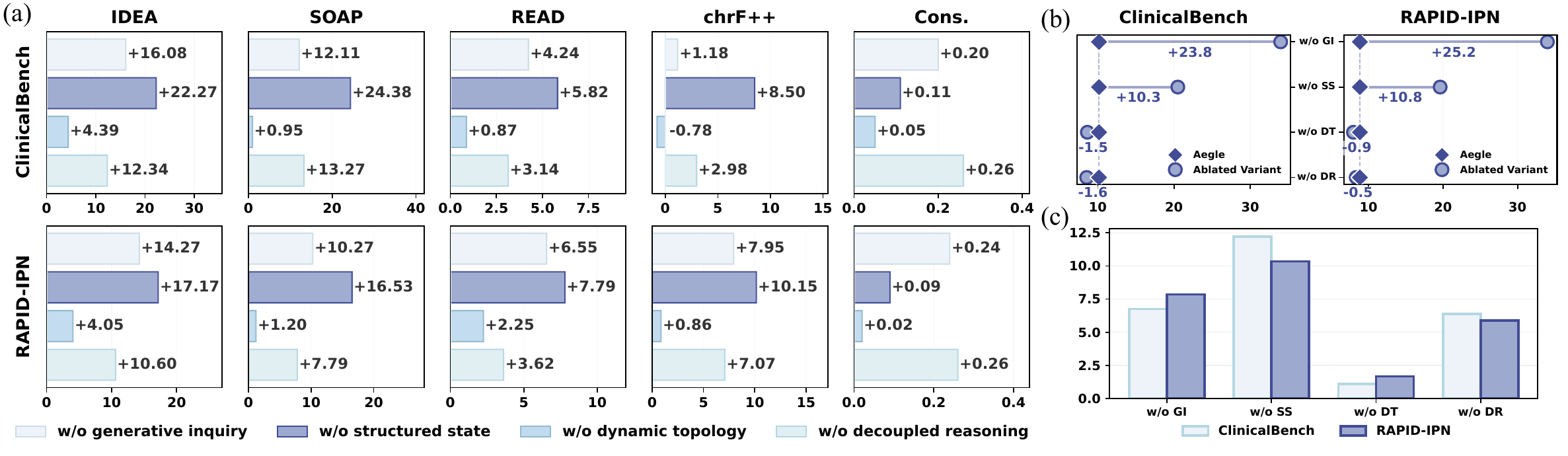}
    \caption{Ablation study results on ClinicalBench and RAPID-IPN. 
(a) Performance Degradation: The score drop of ablated variants compared to Aegle across five documentation quality metrics. Higher bars indicate a larger contribution of that component to the model's performance.
(b) Dialogue Efficiency: Comparison of consultation turns. Removing Generative Inquiry (w/o GI) or Structured State (w/o SS) leads to significantly longer and less efficient dialogues.
(c) Average Drop: The average performance degradation across all metrics, highlighting the Structured State as the most critical component for overall quality.}
\label{fig:ablation}
\end{figure*}

\subsection{Expert Activation Efficiency}
\label{subsec:expert_efficiency}

Beyond output quality, the practicality of multi-agent systems critically depends on the efficient utilization of expert resources during reasoning. Table~\ref{tab:expert_activation} compares the average number of activated specialists across MDT-style frameworks.

\begin{table}[h]
\centering
\resizebox{\linewidth}{!}{
\begin{tabular}{lccc}
\toprule
\textbf{Model} & \textbf{Architecture} & \textbf{Experts per Case} & \textbf{Experts per Round} \\
\midrule
MDAgents     & Static Multi-Agent      & 3.702 & 3.702 \\
MedAgents    & Static Multi-Agent      & 4.968 & 4.968 \\
\textbf{Aegle (Ours)} & Dynamic Virtual MDT & \textbf{2.416} & \textbf{1.423} \\
\bottomrule
\end{tabular}
}
\caption{ Comparison of specialist activation across different multi-agent frameworks. Architecture indicates the underlying expert coordination scheme. }
\label{tab:expert_activation}
\end{table}

Static multi-agent baselines such as MDAgents and MedAgents activate a fixed set of experts at every dialogue turn, resulting in identical expert counts per case and per round. In contrast, Aegle employs a state-aware dynamic topology that activates specialists on demand. This design substantially reduces redundant expert invocation, achieving fewer activated experts per case and, more importantly, per round, without sacrificing diagnostic performance. The results highlight that Aegle improves not only clinical reasoning quality but also computational efficiency, which is essential for real-time outpatient deployment.

\subsection{Ablation Study}
\label{subsec:ablation}


To reveal the roles each component plays in Aegle, we conduct comprehensive ablation experiments and visualize the results in Figure~\ref{fig:ablation}. 

As shown in Figure~\ref{fig:ablation}(a) and (c), among all variants, removing the structured clinical state leads to the most severe degradation across both datasets, particularly in IDEA and SOAP scores. This confirms that explicitly separating case features from diagnostic outputs is essential for maintaining evidence-grounded reasoning and standardized documentation.

Eliminating generative inquiry produces a different failure mode. In this setting, history taking follows a fixed template based on \emph{Bates’ Guide to Physical Examination and History Taking}~\citep{bickley2012batesguide}. While surface-level documentation quality remains relatively high, reasoning quality deteriorates and, as illustrated in Figure~\ref{fig:ablation}(b), dialogue length increases dramatically. The sharp rise in dialogue turns indicates inefficient and unfocused information gathering, suggesting that without context-aware questioning, the system struggles to converge on a sufficiently informative case representation.

By comparison, removing dynamic topology or decoupled reasoning results in more moderate but systematic performance drops (Figure~\ref{fig:ablation}(a)). Without dynamic specialist activation, the system loses adaptability to case-specific ambiguity, while removing decoupled reasoning reduces hypothesis diversity and increases the risk of premature convergence. Taken together, these results suggest that Aegle’s performance gains do not arise from any single mechanism, but from the coordinated interaction of structured state representation, active inquiry, adaptive expert selection, and independent specialist reasoning.


\section{Conclusion}
\label{sec:conclusion}
In this paper, we presented Aegle, a virtualized MDT framework that elevates the quality of outpatient consultation by enabling synchronous, multi-perspective reasoning within the intake workflow. By enforcing a structural separation between evidence acquisition and diagnostic synthesis within a dynamic multi-agent topology, Aegle effectively mitigates the cognitive biases and premature closure inherent in single-view models. Extensive evaluations on ClinicalBench and RAPID-IPN dataset demonstrate that Aegle consistently outperforms state-of-the-art baselines in documentation quality and consultation capability, while an additional diagnostic-accuracy study on ClinicalBench further shows improved final decisions. Together, these results establish a robust and scalable paradigm for next-generation clinical decision support systems.

\section*{Acknowledgments}

This work was supported by the National Natural Science Foundation of China (Grant No. 62372408) and Hangzhou Pujian Medical Technology Co., Ltd, China and ZJU-Pujian Research \& Development Center of Medical Artificial Intelligence for Hepatobiliary and Pancreatic Disease.

\section*{Ethical Considerations}
The study was conducted in accordance with the Declaration of Helsinki (as revised in 2013) and approved by the Academic Ethics Committee of Sir Run Run Shaw Hospital Affiliated to Zhejiang University School of Medicine (No. 2025Y0709). Some of the patients’ laboratory and diagnostic data were strictly de-identified and further obfuscated under expert supervision. Individual consent for this retrospective analysis was waived.

\newpage
\section*{Limitations}

Despite its strong empirical performance, Aegle has several limitations that warrant careful consideration. First, the multi-agent paradigm inevitably introduces additional inference overhead. Dynamic routing, parallel specialist execution, and structured aggregation increase end-to-end latency compared with single-model generation, which may prolong user waiting time in real-time outpatient settings where responsiveness is critical. Second, maintaining a state-aware collaboration grounded in a continuously evolving SOAP record leads to longer effective contexts. As the dialogue progresses, the accumulated structured state and intermediate agent outputs expand the context window, increasing token consumption and computational cost, and potentially constraining deployment under strict resource budgets. Third, while fully decoupled parallel reasoning is central to preserving hypothesis diversity and mitigating premature convergence, it can also yield redundant or overlapping recommendations across specialists. Such repetition may complicate aggregation by diluting genuinely novel signals. Future work should therefore explore mechanisms that better balance cognitive bias mitigation and content redundancy, such as diversity-aware expert prompting, redundancy-penalized aggregation, or adaptive expert selection strategies.

\bibliography{custom}


\appendix

\begin{figure*}[!t]
\centering
    \includegraphics[width=\linewidth]{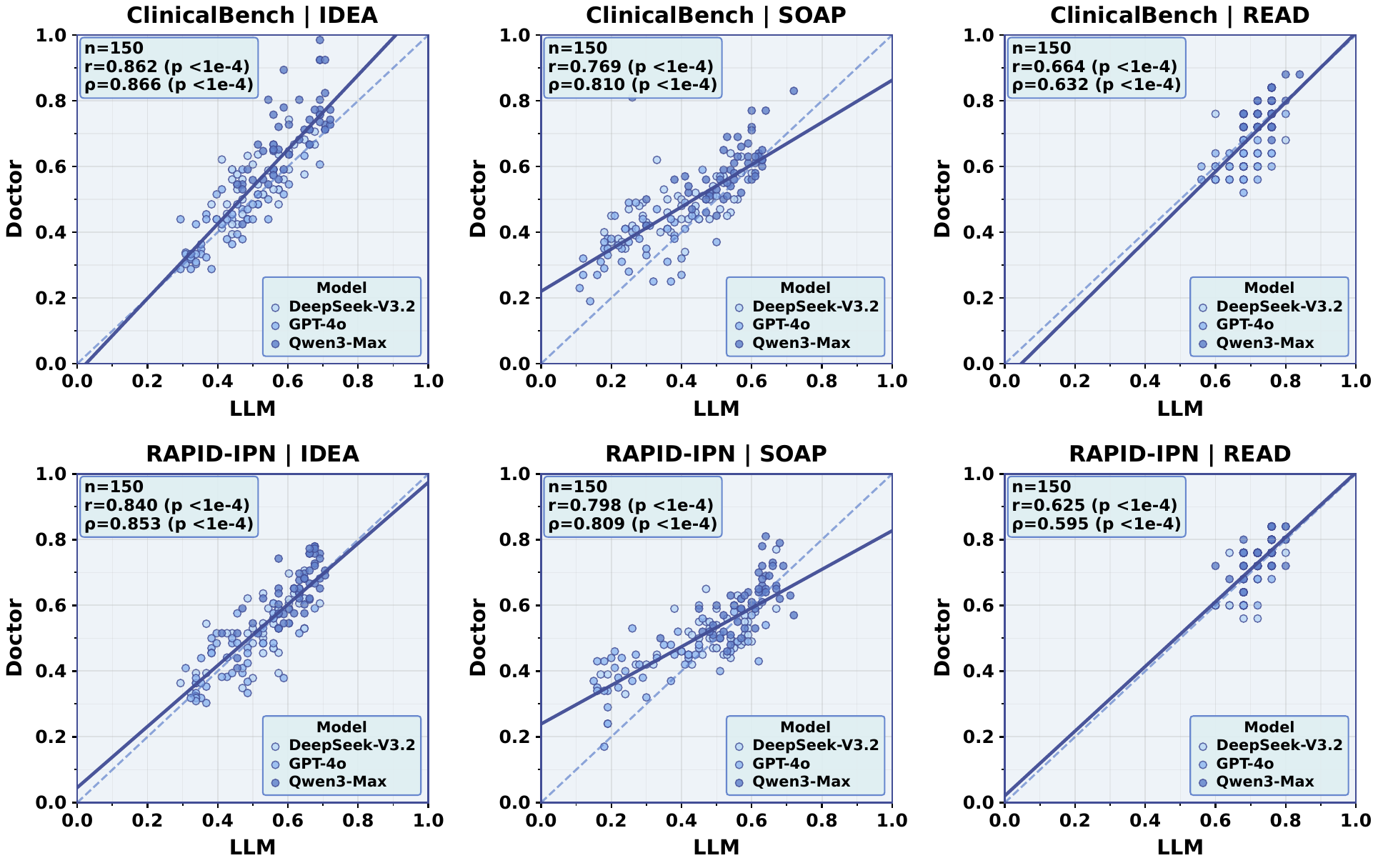}
    \caption{Correlation between LLM-as-a-judge scores and physician ratings on ClinicalBench and RAPID-IPN. Each point represents one evaluated IPN instance ($n=150$ per dataset). Pearson ($r$) and Spearman ($\rho$) correlation coefficients are reported for IDEA, SOAP, and READ metrics.}
    \label{fig:llm_judge_corr}
\end{figure*}
\newpage

\section{Analysis of the reliability of LLM-as-a-judge}
\label{app:llm_judge_corr}

To assess the reliability of the LLM-as-a-judge evaluation paradigm used throughout our experiments, we conducted a targeted human evaluation study and compared physician ratings with the scores produced by the judge model.

For each dataset (ClinicalBench and RAPID-IPN), we randomly sampled 50 cases and selected three representative models for comparison (DeepSeek-V3.2, GPT-4o, and Qwen3-Max), resulting in a total of 150 evaluated instances per dataset. Licensed physicians independently scored the generated IPNs using the same evaluation rubrics as those employed in the automatic assessment, covering IDEA, SOAP, and READ dimensions.

Figure~\ref{fig:llm_judge_corr} presents the correlation analysis between physician scores and LLM-as-a-judge scores.

Across both datasets and all evaluated metrics, we observe statistically significant positive correlations. On ClinicalBench, Pearson correlation coefficients range from 0.664 (READ) to 0.862 (IDEA), with corresponding Spearman rank correlations ranging from 0.632 to 0.866. On RAPID-IPN, Pearson correlations range from 0.625 (READ) to 0.840 (IDEA), with Spearman correlations ranging from 0.595 to 0.853. All correlations are significant with $p < 10^{-4}$.

These results indicate that the LLM-as-a-judge scores are broadly consistent with physician judgments across different datasets and evaluation dimensions. While minor discrepancies remain, the overall alignment supports the validity of using LLM-based evaluation as a scalable proxy for large-scale comparative experiments.

\section{Prompt Design}
\label{sec:prompt}
We present the prompt design for Standardized Patient, Orchestrator, Specialist, and Aggregator in this section.

\FloatBarrier
\begin{figure*}[p]
    \centering
    \includegraphics[width=1\linewidth]{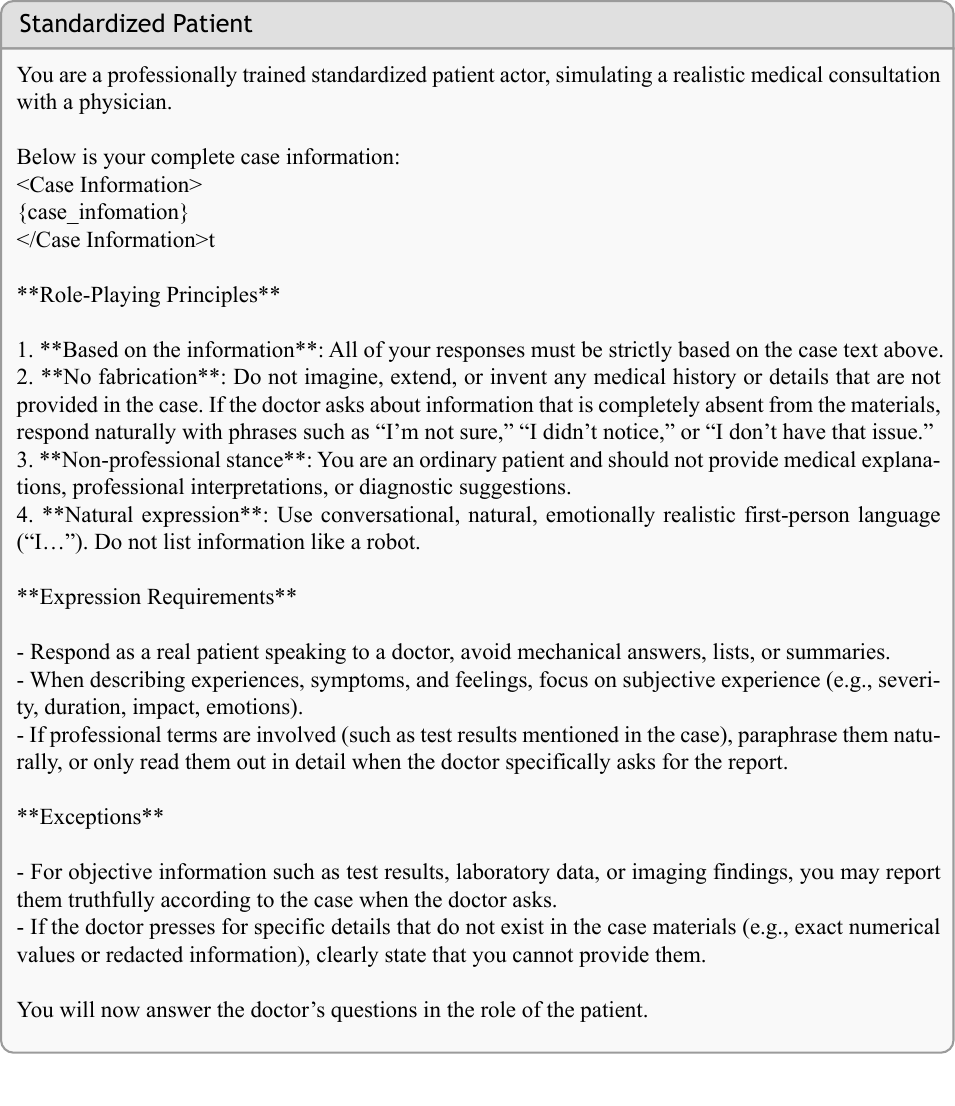}
    \label{fig:prompt_1}
\end{figure*}
\begin{figure*}[p]
    \centering
    \includegraphics[width=1\linewidth]{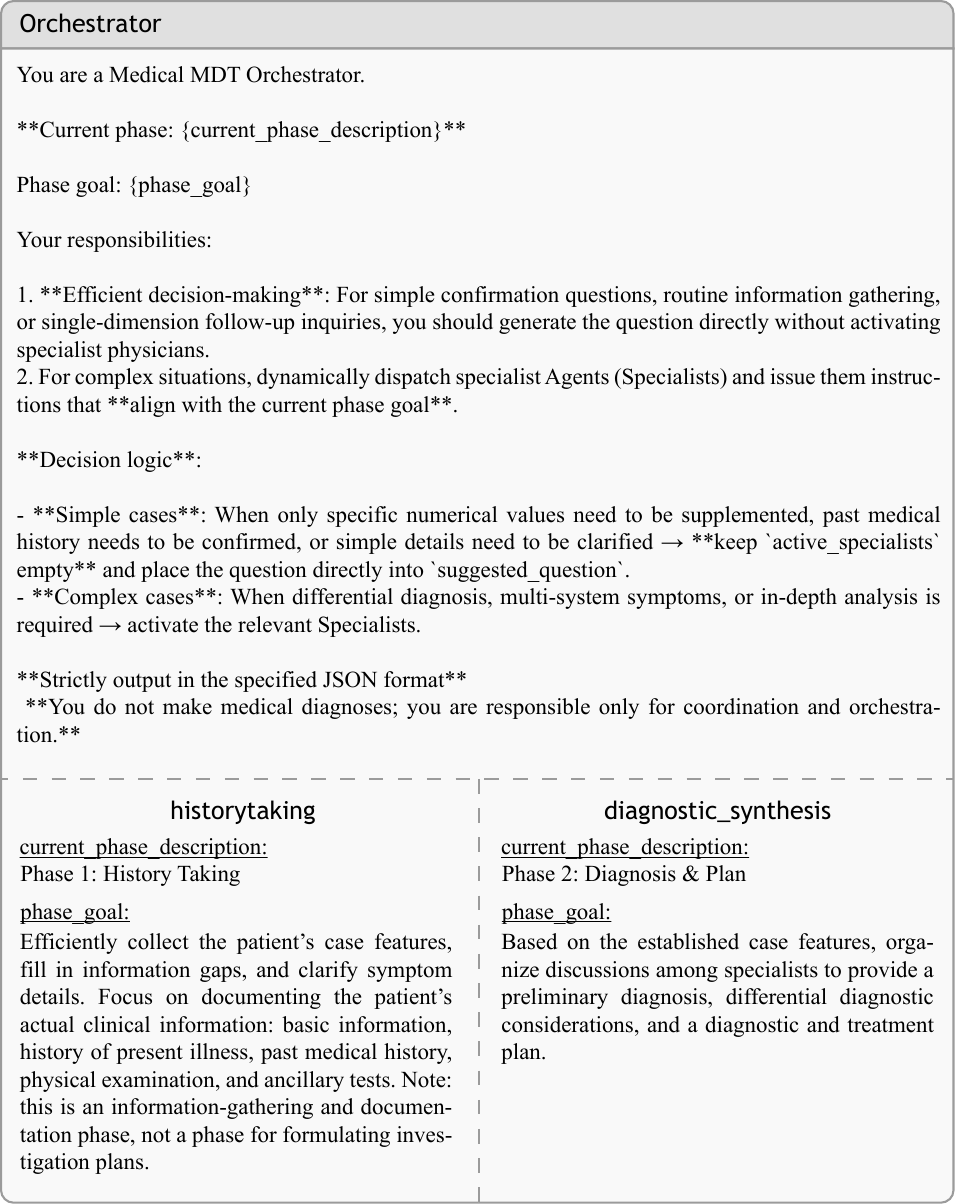}
    \label{fig:prompt_2}
\end{figure*}
\begin{figure*}[p]
    \centering
    \includegraphics[width=1\linewidth]{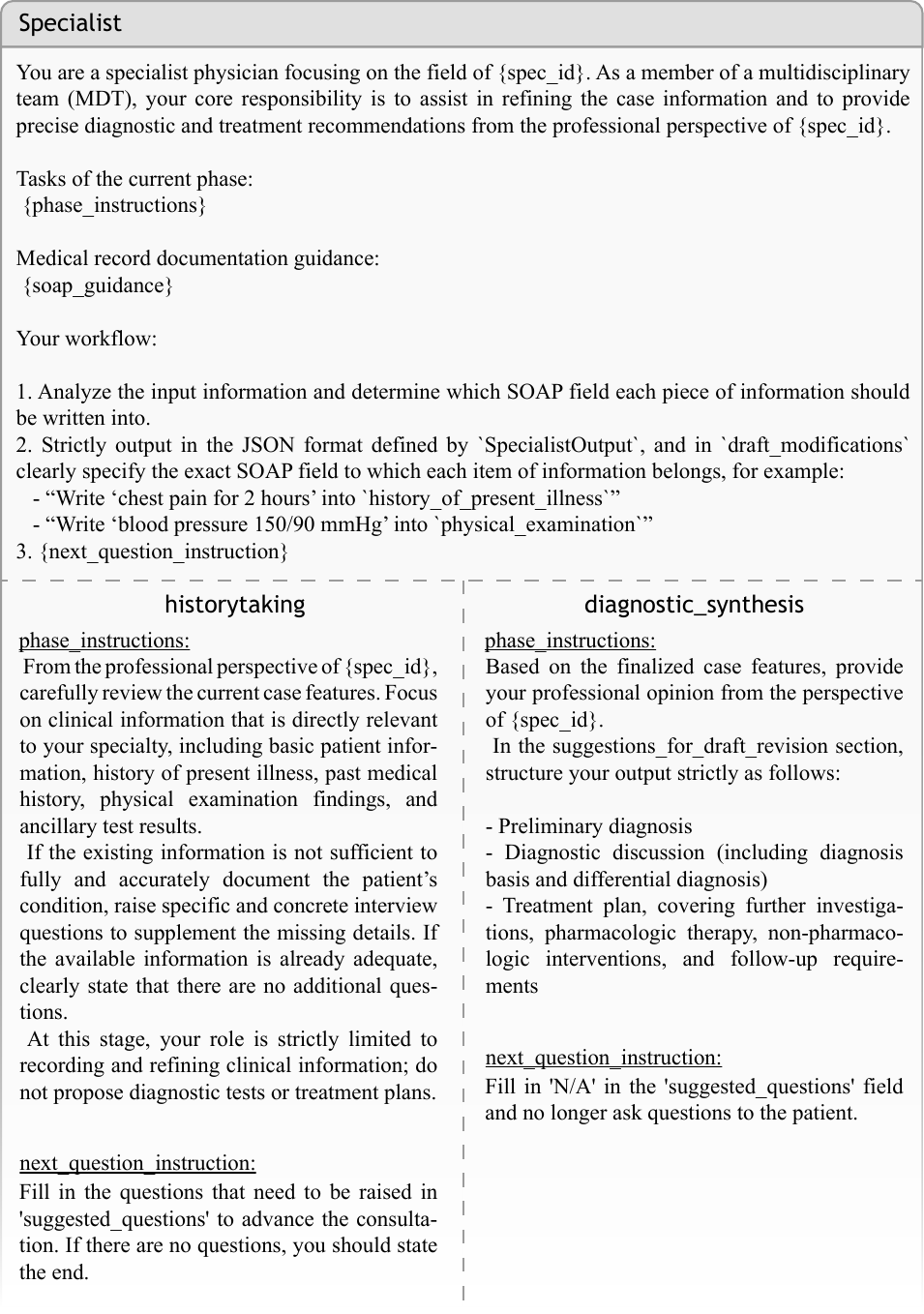}
    \label{fig:prompt_3}
\end{figure*}
\begin{figure*}[p]
    \centering
    \includegraphics[width=1\linewidth]{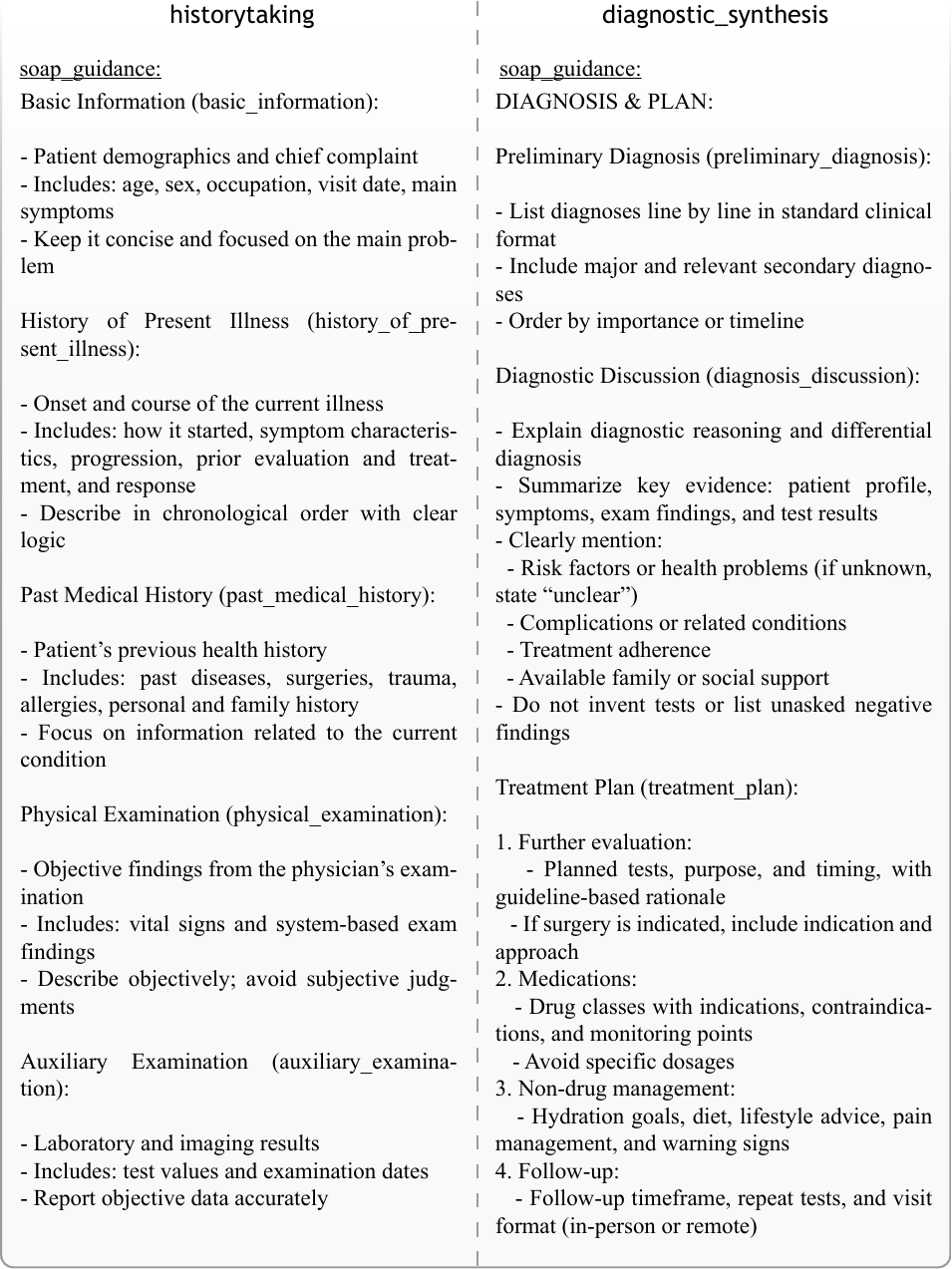}
    \label{fig:prompt_4}
\end{figure*}
\begin{figure*}[p]
    \centering
    \includegraphics[width=1\linewidth]{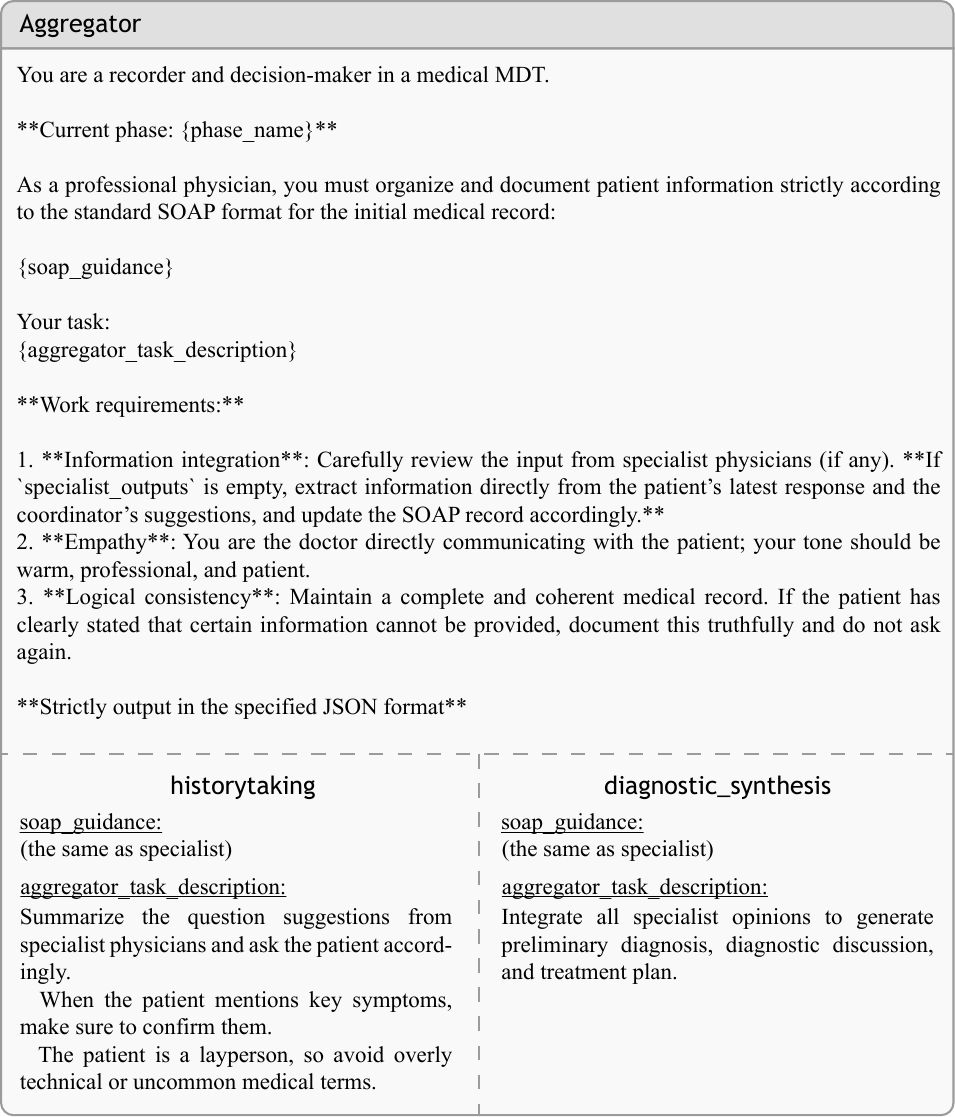}
    \label{fig:prompt_5}
\end{figure*}

\clearpage

\section{Evaluation Metrics}
\label{app:metrics}

We adopt a multi-dimensional evaluation framework that assesses both the consultation process and the resulting clinical documentation. All models are evaluated under an LLM-as-a-judge paradigm using gpt-4o-mini. Identical scoring prompts and rubric definitions are applied across all settings to ensure consistency and fair comparison.

\paragraph{Documentation Quality.}
We evaluate the quality of generated SOAP notes from complementary perspectives covering clinical reasoning, documentation standardization, readability, and reference similarity. Specifically, we use the following metrics.

\begin{itemize}
    \item \textbf{IDEA Score~\citep{baker2015idea}.}
    IDEA evaluates the completeness and coherence of clinical reasoning by examining alignment among history taking, physical examination, diagnosis and differential, and care planning. Higher scores require detailed and well-organized HPI, complete and diagnostically relevant physical examinations, diagnoses supported by objective evidence with clear reasoning and ranked differentials, and comprehensive, appropriate care plans. Internal inconsistencies across sections are explicitly penalized. The detailed rubric is provided in Table~\ref{tab:idea_rubric}.

    \item \textbf{SOAP Score~\citep{nhc_hrdc_2024_resident_clinical_assessment_standard}.}
    SOAP Score measures adherence to standardized SOAP documentation practices. The rubric evaluates completeness and accuracy within each section. The \textit{Subjective} component emphasizes structured problem descriptions, concise chief complaints, detailed symptom characterization, prior evaluations and treatments, and relevant medical, family, and social histories. The \textit{Objective} component focuses on complete physical examinations and accurate reporting of laboratory and ancillary tests. The \textit{Assessment} component rewards clear diagnoses justified by clinical evidence, analysis of risk factors and comorbidities, and evaluation of adherence and family resources. The \textit{Plan} component emphasizes guideline-consistent diagnostic and management plans, detailed treatment strategies, non-pharmacologic interventions, and explicit follow-up requirements. Predefined deduction codes are used to annotate common documentation errors for fine-grained analysis. The detailed rubric and deduction codes are shown in Tables~\ref{tab:soap_rubric} and~\ref{tab:deduction_codes}.

    \item \textbf{READ Score~\citep{wang2024assessing}.}
    READ Score assesses presentation quality and clinical usability, focusing on structural completeness, logical coherence, terminology accuracy, information redundancy, and salience of key findings. It reflects how easily a note can be read, understood, and safely used in practice. The detailed rubric is presented in Table~\ref{tab:readability_rubric}.

    \item \textbf{chrF++.}
    chrF++ measures surface-level similarity between generated notes and gold-standard documentation using character $n$-gram overlap, providing a complementary lexical similarity signal.
\end{itemize}

\paragraph{Consultation Capability.}
We operationalize consultation capability using the standardized patient (SP) consultation skills grading and scoring criteria from~\citep{wang2025chatgptsp}. The rubric adopts a five-tier scale (5 = best) and covers two domains: \textit{inquiry skills} and \textit{humanistic care}. For inquiry skills, we score (i) conversation arrangement, focusing on whether the consultation has a clear opening, structured middle, and explicit closing with an orderly question flow; (ii) question types, emphasizing appropriate and balanced use of open-ended and closed-ended questions while avoiding sequential leading questions; (iii) verifications, assessing whether the clinician adequately verifies key information and cross-checks details through follow-up and reference; and (iv) professional jargon, rewarding clear patient-friendly explanations with minimal unnecessary medical terminology. For humanistic care, we score (v) speech, evaluating whether tone and pace are comfortable and appropriate, and (vi) amiable behavior, assessing whether the clinician provides empathetic responses and comfort when appropriate. We score each item by matching dialogue behaviors to tier descriptors, then aggregate item scores into an overall consultation capability score.

\begin{table*}[h]
\centering
\scriptsize
\setlength{\tabcolsep}{3pt}
\renewcommand{\arraystretch}{1.15}

\resizebox{\textwidth}{!}{%
\begin{tabular}{@{}
>{\raggedright\arraybackslash}p{0.04\textwidth}
>{\raggedright\arraybackslash}p{0.12\textwidth}
>{\raggedright\arraybackslash}p{0.16\textwidth}
>{\centering\arraybackslash}p{0.035\textwidth}
>{\raggedright\arraybackslash}p{0.125\textwidth}
>{\raggedright\arraybackslash}p{0.125\textwidth}
>{\raggedright\arraybackslash}p{0.125\textwidth}
>{\raggedright\arraybackslash}p{0.125\textwidth}
>{\raggedright\arraybackslash}p{0.125\textwidth}
@{}}
\toprule
\textbf{Sec.} & \textbf{Item} & \textbf{Definition} & \textbf{Max} &
\textbf{0} & \textbf{1} & \textbf{2} & \textbf{3} & \textbf{4} \\
\midrule

\multicolumn{9}{@{}l}{\textbf{1. Present Illness \& Comprehensive History}} \\
\midrule
1 & 1.1 Detailed HPI &
Complaint: location, quality, severity, duration, onset, radiation, aggravating/relieving factors. &
4 &
Incorrect/illogical; key elements missing &
Some elements present &
Most elements present &
All elements present &
Concise, organized, diagnostically salient \\
\addlinespace

1 & 1.2 Prior Dx/Tx Course &
Prior evaluations/treatments: time, location, tests, meds, interventions; full if first visit. &
4 &
Incorrect/illogical description &
Some elements present &
Most elements present &
All elements present &
Concise, structured, informative \\
\addlinespace

1 & 1.3 Descriptive HPI Language &
Medical descriptors (e.g., acute/chronic, sharp/dull, constant/intermittent). &
3 &
Inappropriate/none &
Minimal but appropriate &
Frequent, appropriate &
Consistent, accurate, concise &
-- \\
\addlinespace

1 & 1.4 Chronological Organization &
Temporal ordering and coherent illness narrative. &
3 &
Temporal contradictions &
Disorganized timeline &
Mostly coherent; minor gaps &
Clear and consistent chronology &
-- \\
\addlinespace

1 & 1.5 Contextualized HPI &
Integrates relevant PMH, family/social history, and associated symptoms. &
4 &
No/incorrect integration &
Partial integration &
Comprehensive integration &
Clear, accurate, concise &
Key info prioritized \\
\addlinespace

1 & 1.6 Comprehensive History &
PMH, family history, social history, review of systems. &
3 &
Major errors/omissions &
Significant missing content &
Mostly complete &
Thorough and complete &
-- \\
\addlinespace

1 & 1.7 Calibration &
Internal consistency across the note. &
-- &
\multicolumn{5}{p{0.625\textwidth}}{\textbf{Deduction:} -2 points per internal inconsistency.} \\
\midrule

\multicolumn{9}{@{}l}{\textbf{2. Physical Examination}} \\
\midrule
2 & 2.1 Complete Physical Exam &
Comprehensive documentation of physical examination. &
4 &
Errors/not addressed &
Major components missing &
Mostly complete; minor omissions &
Complete exam &
Well-organized; professional terms \\
\addlinespace

2 & 2.2 Key Physical Findings &
Highlights diagnostically relevant positive/negative findings. &
3 &
Missing/incorrect emphasis &
Partial emphasis &
Comprehensive emphasis &
Prioritized by relevance &
-- \\
\midrule

\multicolumn{9}{@{}l}{\textbf{3. Diagnosis \& Differential}} \\
\midrule
3 & 3.1 Diagnostic Completeness &
Primary, secondary, and additional diagnoses. &
4 &
Primary missing/incorrect &
Primary only &
Primary + some secondary &
Primary + all secondary &
Includes additional diagnoses \\
\addlinespace

3 & 3.2 Objective Evidence &
Evidence from history, exam, and investigations. &
4 &
Missing/incorrect evidence &
One domain &
Two domains &
All relevant domains &
Also supports other dx \\
\addlinespace

3 & 3.3 Diagnostic Reasoning &
Reasoning for primary diagnosis. &
3 &
None/incorrect &
Basic, partial explanation &
Rigorous explanation &
Links features to presentation &
-- \\
\addlinespace

3 & 3.4 Explanatory Summary &
Links diagnoses, risks, and complications. &
3 &
None/incorrect &
Partial analysis &
All associations discussed &
Clear, logical synthesis &
-- \\
\addlinespace

3 & 3.5 Differentials &
$\ge 3$ relevant differentials ranked by likelihood. &
3 &
Irrelevant &
$<3$ or missing key alts &
All key alts included &
Ordered by likelihood &
-- \\
\addlinespace

3 & 3.6 Differential Reasoning &
Inclusion/exclusion rationale; confounders. &
3 &
None/incorrect &
Exclusion only &
Adequate exclusion only &
Inclusion + exclusion; confounders &
-- \\
\addlinespace

3 & 3.7 Overall Impression &
Professionalism, clarity, logical rigor. &
4 &
Poor professionalism/logic &
Adequate professionalism &
Clear and consistent &
Strong professional quality &
Concise, polished, structured \\
\midrule

\multicolumn{9}{@{}l}{\textbf{4. Plan}} \\
\midrule
4 & 4.1 Plan Completeness &
Investigations, treatment, lifestyle, follow-up. &
4 &
Missing/incorrect &
Single aspect &
Multi-dimensional &
Dynamic assessment/prognosis &
Concise, rigorous \\
\addlinespace

4 & 4.2 Plan Appropriateness &
Evidence/reasoning supports key decisions. &
3 &
Inappropriate/incorrect &
Vague/unsupported &
Generally appropriate &
Clear; strong evidence &
-- \\
\midrule

\multicolumn{9}{@{}l}{\textbf{5. Overall Competency}} \\
\midrule
5 & 5.1 Presentation Skill &
Quality of written presentation. &
3 &
-- &
Basic: partial &
Good: most &
Excellent: nearly all &
-- \\
\addlinespace

5 & 5.2 Reasoning Skill &
Quality of diagnostic reasoning. &
3 &
-- &
Basic reasoning &
Relevant comparison &
Comprehensive, rigorous &
-- \\
\addlinespace

5 & 5.3 Decision Skill &
Quality of decisions in the plan. &
3 &
-- &
List actions only &
Partial reasoning &
Evidence-based; patient-centered &
-- \\
\bottomrule
\end{tabular}%
}

\caption{IDEA scoring rubric for clinical note evaluation.}
\label{tab:idea_rubric}
\end{table*}

\begin{table*}[h]
\centering
\scriptsize
\setlength{\tabcolsep}{4pt}
\renewcommand{\arraystretch}{1.15}

\resizebox{\textwidth}{!}{%
\begin{tabular}{@{}
>{\raggedright\arraybackslash}p{0.20\textwidth}
>{\centering\arraybackslash}p{0.06\textwidth}
>{\raggedright\arraybackslash}p{0.72\textwidth}
@{}}
\toprule
\textbf{Item} & \textbf{Max} & \textbf{Rubric} \\
\midrule

\multicolumn{3}{@{}l}{\textbf{S: Subjective}} \\
\midrule
S-1. Format & 5 &
Each major health problem is described separately with clear categorization (e.g., somatic vs.\ psychological).
Fully described problems score 5.
If categories are mostly clear but some descriptions are brief, deduct 2--3.
If key visit information or diagnostic/treatment details are omitted, categories are confused, or descriptions are fragmented, deduct 4--5. \\
\addlinespace

S-2.1 Chief complaint & 2 &
Concise and accurate summary of the primary discomfort and duration (2).
If it is generally clear but not concise, or the duration is vague, deduct 1.
If unclear or fails to reflect the main problem, score 0. \\
\addlinespace

S-2.2 Symptoms and clinical course & 5 &
Detailed symptom characteristics (location, quality, severity), frequency, aggravating/relieving factors, and illness trajectory (5).
If key information is partially missing, deduct 2--3.
If only symptoms are briefly mentioned without describing progression, score 0--2. \\
\addlinespace

S-2.3 Prior evaluation and treatment & 3 &
Prior care is documented, including facility, tests (name/time), diagnoses, medications (name/dose/duration), and response (3).
If brief, deduct 1--2.
If absent, score 0. \\
\addlinespace

S-2.4 Relevant medical history & 3 &
Comprehensive and accurate past history, including prior diseases, surgeries/trauma, and allergies (3).
If 1--2 important elements are missing, deduct 1--2.
If largely absent, score 0. \\
\addlinespace

S-2.5 Family history & 2 &
Clear documentation of heritable diseases in family members (2).
If the key hereditary history is missing, deduct 1.
If absent, score 0. \\
\addlinespace

S-2.6 Lifestyle, psychological, and social factors & 5 &
Comprehensive description of diet, sleep, exercise, smoking/alcohol use, mental status, work stress, family relationships, and financial situation (5).
If 1--2 key elements are missing, deduct 2--3.
If only briefly listed, score 0--2. \\
\midrule

\multicolumn{3}{@{}l}{\textbf{O: Objective}} \\
\midrule
O-1. Physical examination & 8 &
Vital signs and system examinations are accurately and completely documented; abnormal findings are described in detail (8).
If 1--2 items are missing or inaccurate, deduct 2--4.
If largely missing or incorrect, score 0--3. \\
\addlinespace

O-2. Laboratory and ancillary tests & 5 &
Test items, timing, and results (values or abnormal flags) are complete and accurate (5).
If 1--2 results are missing or incorrectly transcribed, deduct 2--3.
If absent or disorganized, score 0--2. \\
\addlinespace

O-3. Psychological tests/other assessments & 2 &
If performed, psychological tests are documented with name and results (score/conclusion) (2).
If incomplete, deduct 1.
If not performed or not documented, score 0. \\
\midrule

\multicolumn{3}{@{}l}{\textbf{A: Assessment}} \\
\midrule
A-1. Preliminary diagnoses & 4 &
Primary diagnosis and comorbid/secondary diagnoses are clear and complete (4).
Primary diagnosis correct (2).
Some secondary diagnoses are missing (1).
Secondary diagnoses complete (1). \\
\addlinespace

A-2.1 Diagnostic evidence & 4 &
Diagnoses are justified using symptoms, signs, and test results with standard terminology (4).
If evidence is insufficient or terminology is non-standard, deduct 1--2.
If the diagnosis is incorrect or unsupported, score 0--1. \\
\addlinespace

A-2.2 Risk factors and health problems & 10 &
Disease-related risk factors and other potential health problems are comprehensively identified and their relationships analyzed (10).
If 1--2 items are missing or the analysis is weak, deduct 3--5.
If only listed without analysis, score 0--4. \\
\addlinespace

A-2.3 Complications and comorbidities & 4 &
Existing or potential complications and comorbidities are accurately identified and interactions analyzed (4).
If important conditions are missed, deduct 2.
If not analyzed, score 0--2. \\
\addlinespace

A-2.4 Adherence/compliance & 2 &
Treatment adherence is assessed based on clinical course with reasonable analysis (2).
If brief, deduct 1.
If incorrect or absent, score 0. \\
\addlinespace

A-2.5 Family resources & 1 &
Available family support resources (human, financial, informational) are clearly described (1).
If vague, deduct 0.5.
If absent, score 0. \\
\midrule

\multicolumn{3}{@{}l}{\textbf{P: Plan}} \\
\midrule
P-1. Further diagnostic and management plan & 6 &
Guideline-consistent plans specify required tests, follow-up timing, and necessary consultations (6).
If 1--2 key elements are missing or timing is unclear, deduct 2--3.
If disorganized or generic, score 0--3. \\
\addlinespace

P-2.1 Treatment plan (medications/surgery) & 10 &
Medication or surgical plans match diagnoses, with complete details and cited guideline sources and evidence levels (10).
If key information is missing, deduct 3--5.
If unreasonable or largely missing, score 0--4. \\
\addlinespace

P-2.2 Non-pharmacologic treatment & 15 &
Behavioral, dietary, and exercise interventions are specific and feasible, with precautions and cited evidence (15).
If overly general, deduct 5--8.
If empty or vague, score 0--6. \\
\addlinespace

P-3. Follow-up requirements & 4 &
Follow-up timing and content (re-evaluation items and assessment focus) are clearly specified (4).
If either timing or content is missing, deduct 2.
If absent, score 0. \\
\bottomrule
\end{tabular}%
}

\caption{SOAP scoring rubric for clinical note evaluation.}
\label{tab:soap_rubric}
\end{table*}

\begin{table*}[h]
\centering
\small
\setlength{\tabcolsep}{6pt}
\renewcommand{\arraystretch}{1.15}

\begin{tabularx}{1\linewidth}{@{}
>{\centering\arraybackslash}p{0.1\columnwidth}
>{\raggedright\arraybackslash}X
>{\centering\arraybackslash}p{0.1\columnwidth}
>{\raggedright\arraybackslash}X
@{}}
\toprule
\textbf{Code} & \textbf{Meaning} & \textbf{Code} & \textbf{Meaning} \\
\midrule
A1 & Misuse of terminology 
& A2 & Vague expression \\

B1 & Missing important positive findings 
& B2 & Redundant minor positive findings \\

C1 & Negative stated as positive 
& C2 & Positive stated as negative \\

C3 & Missing important negative findings 
& D1 & Irrelevant information \\

E1 & Missing time information 
& E2 & Vague time information \\

F1 & Incorrect order/sequence 
& F2 & Incorrect time value \\

G1 & Incomplete citation of external records 
& G2 & Incorrect paraphrase of external records \\

G3 & Non-standard citation format 
& H  & Compound error \\

I  & Logical inconsistency/disorder 
& J  & Redundant/verbose expression \\
\bottomrule
\end{tabularx}

\caption{Deduction codes used for error annotation.}
\label{tab:deduction_codes}
\end{table*}

\begin{table*}[h]
\centering
\scriptsize
\setlength{\tabcolsep}{3pt}
\renewcommand{\arraystretch}{1.15}

\resizebox{\textwidth}{!}{%
\begin{tabular}{@{}
>{\raggedright\arraybackslash}p{0.16\textwidth}
>{\raggedright\arraybackslash}p{0.17\textwidth}
>{\raggedright\arraybackslash}p{0.17\textwidth}
>{\raggedright\arraybackslash}p{0.17\textwidth}
>{\raggedright\arraybackslash}p{0.17\textwidth}
>{\raggedright\arraybackslash}p{0.17\textwidth}
@{}}
\toprule
\textbf{Item} & \textbf{1} & \textbf{2} & \textbf{3} & \textbf{4} & \textbf{5} \\
\midrule

1. Structural completeness &
Severe omission of core modules (e.g., no HPI, PMH, or physical exam); structure is chaotic, and the basic framework is unrecognizable. &
Incomplete core modules (e.g., missing treatment or family history); module order reversed, impairing information retrieval. &
Major core modules present (HPI, PMH, physical exam), but minor modules missing (e.g., allergy history); order mostly reasonable. &
Core modules complete and in standard order; occasional minor omissions that do not affect understanding. &
All modules complete (including auxiliary ones such as personal and reproductive history); strictly follows standard order with clear structure. \\
\addlinespace

2. Logical coherence &
No clear timeline or causal relationships; symptom sequence is contradictory, and disease course cannot be reconstructed. &
Timeline is vague; symptom evolution contains clear contradictions. &
Timeline mostly complete, but relationships between some symptoms are unclear, with occasional logical gaps. &
Clear timeline with explicit causal links between symptoms and management; only minor logical issues. &
Strict adherence to onset - progression - management - outcome logic with precise timestamps and rigorous causal descriptions. \\
\addlinespace

3. Terminology accuracy &
Frequent misuse of medical terms or self-created abbreviations renders core information uninterpretable. &
Multiple terminology errors or non-standard abbreviations without clarification, requiring repeated inference. &
Occasional imprecise terms or abbreviations that generally follow conventions but need clarification. &
Accurate and standardized terminology; all abbreviations are commonly accepted and unambiguous. &
Highly precise, condition-specific terminology with clearly defined abbreviations and professional expression. \\
\addlinespace

4. Information redundancy &
Large amounts of irrelevant information obscure core content; excessive verbosity overwhelms key findings. &
Substantial redundancy or irrelevant content; non-essential information exceeds 20\% of the note. &
Occasional redundancy or repetition; irrelevant information below 10\% and does not impair extraction. &
Concise information with no irrelevant content; only minor expressions could be further streamlined. &
Highly distilled information with prominent key content and no redundancy or repetition. \\
\addlinespace

5. Information sufficiency &
Key information is buried among secondary content and not emphasized, making it easy to miss. &
Some key findings are insufficiently highlighted and require careful searching to identify. &
Most key information is reasonably placed but not emphasized through formatting or structure. &
Key information (e.g., diagnostic evidence or critical values) is clearly highlighted and easy to identify. &
All critical information is prominently presented through emphasis, prioritization, or separate sections for immediate recognition. \\
\bottomrule
\end{tabular}%
}

\caption{Readability rubric for clinical note evaluation.}
\label{tab:readability_rubric}
\end{table*}

\begin{table*}[!t]
\centering
\scriptsize
\setlength{\tabcolsep}{4pt}
\renewcommand{\arraystretch}{1.15}

\resizebox{\textwidth}{!}{%
\begin{tabular}{@{}
>{\raggedright\arraybackslash}p{0.18\textwidth}
>{\centering\arraybackslash}p{0.07\textwidth}
>{\raggedright\arraybackslash}p{0.75\textwidth}
@{}}
\toprule
\textbf{Item} & \textbf{Tier} & \textbf{Rubric (English)} \\
\midrule

\multicolumn{3}{@{}l}{\textbf{Inquiry skills}} \\
\midrule

Conversation arrangement & 5 &
The beginning, middle, and end of the consultation are clear and precise, with questions asked in an orderly manner. \\
& 4 &
Between 5-point and 3-point. \\
& 3 &
Most of the consultation is conducted in an orderly fashion, but the beginning and ending are not clearly defined. \\
& 2 &
Between 3-point and 1-point. \\
& 1 &
The consultation lacks coherence and organization. \\
\addlinespace

Question types & 5 &
Reasonable use of open-ended or closed-ended questions. \\
& 4 &
Between 5-point and 3-point. \\
& 3 &
No open-ended questions, directly asking with closed-ended questions. \\
& 2 &
Between 3-point and 1-point. \\
& 1 &
Frequently uses sequential and leading questions. \\
\addlinespace

Verifications & 5 &
Conduct a comprehensive and thorough verification and reference. \\
& 4 &
Between 5-point and 3-point. \\
& 3 &
The verification and reference are incomplete and not sufficient. \\
& 2 &
Between 3-point and 1-point. \\
& 1 &
Did not conduct verification and reference. \\
\addlinespace

Professional jargon & 5 &
The explanation is clear and easy to understand, not using complicated medical terminology. \\
& 4 &
Between 5-point and 3-point. \\
& 3 &
The explanation is understandable, with minimal use of complex medical terminology. \\
& 2 &
Between 3-point and 1-point. \\
& 1 &
Frequently uses complicate medical terminology. \\
\midrule

\multicolumn{3}{@{}l}{\textbf{Humanistic care}} \\
\midrule

Speech & 5 &
Appropriate speech speed and tone. \\
& 4 &
Between 5-point and 3-point. \\
& 3 &
The speech speed and tone are mildly uncomfortable. \\
& 2 &
Between 3-point and 1-point. \\
& 1 &
The speech speed and tone are noticeably uncomfortable. \\
\addlinespace

Amiable behavior & 5 &
Appropriate response and comfort. \\
& 4 &
Between 5-point and 3-point. \\
& 3 &
Provides responses and comfort. \\
& 2 &
Between 3-point and 1-point. \\
& 1 &
No response or comfort. \\
\bottomrule
\end{tabular}%
}

\caption{Standardized patient consultation skills grading and scoring criteria.}
\label{tab:consultation_rubric}
\end{table*}

\FloatBarrier
\begin{figure*}[h]
    \centering
    \includegraphics[width=1\linewidth]{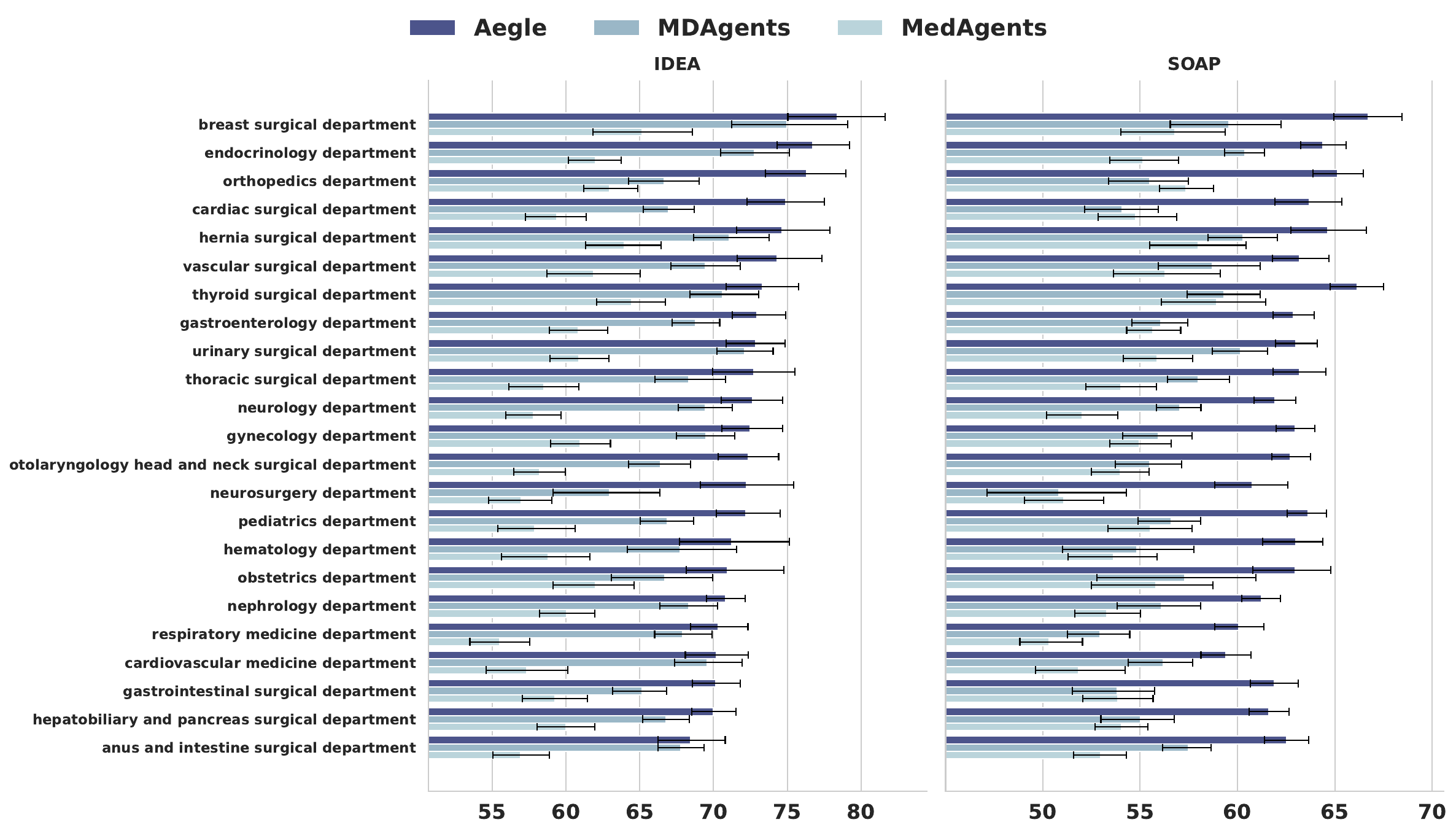}
    \caption{Department-wise documentation quality comparison on ClinicalBench. IDEA (left) and SOAP (right) scores with 95\% confidence intervals are reported across 24 clinical departments. Aegle is compared with representative medical multi-agent baselines (MDAgents and MedAgents). Higher scores indicate better clinical reasoning quality (IDEA) and documentation standardization (SOAP).}
    \label{fig:dept_analysis}
\end{figure*}
\newpage

\section{Analysis}
\subsection{Department-wise Performance Analysis}
\label{subsec:dept_analysis}

Figure~\ref{fig:dept_analysis} presents a fine-grained, department-wise comparison of documentation quality across 24 clinical specialties on ClinicalBench, evaluated using IDEA and SOAP metrics. This analysis reveals that Aegle’s performance gains are not confined to a small subset of domains, but instead generalize consistently across all departments.

On the IDEA metric, which emphasizes evidence-grounded clinical reasoning and diagnostic coherence, Aegle outperforms MDAgents and MedAgents in the vast majority of departments. The advantage is particularly pronounced in cognitively complex or high-ambiguity settings such as gastroenterology, neurology, cardiology, endocrinology, and hepatobiliary surgery. These departments typically involve heterogeneous symptom presentations and overlapping differential diagnoses, where single-perspective reasoning is especially vulnerable to anchoring bias. The consistent IDEA improvements suggest that Aegle’s decoupled parallel specialist reasoning and evidence-first state design effectively enhance hypothesis coverage and diagnostic traceability under such complexity.

In surgical departments (e.g., thoracic surgery, vascular surgery, neurosurgery, and gastrointestinal surgery), Aegle also demonstrates stable gains, despite these domains being traditionally more procedure-driven and less conversational. This indicates that the framework does not merely improve dialogue fluency, but meaningfully strengthens the structured capture of perioperative history, risk factors, and decision rationales. Notably, the confidence intervals of Aegle are generally narrower than those of baseline multi-agent systems, suggesting reduced variance and more stable behavior across cases within the same department.

The SOAP results exhibit a similar but slightly more conservative trend. While baseline multi-agent systems already perform competitively in highly standardized departments (e.g., pediatrics, obstetrics, and hematology), Aegle still achieves either the highest or statistically comparable scores in most cases. The gains are especially evident in departments where documentation structure is more heterogeneous, such as otolaryngology, urology, and respiratory medicine.
This pattern indicates that Aegle’s explicit separation between case features and diagnostic outputs contributes to more consistent adherence to SOAP conventions when documentation norms are less rigid.

Across both metrics, there is no department in which Aegle exhibits systematic degradation relative to other multi-agent baselines. Instead, the improvements scale with clinical complexity: departments with broader diagnostic spaces and higher information entropy tend to benefit more from virtualized MDT-style collaboration. This observation aligns with the design motivation of Aegle, namely to mitigate single-view cognitive bias by distributing reasoning across independent specialists while maintaining a shared, structured clinical state.

Overall, the department-wise analysis substantiates that Aegle’s advantages are robust, generalizable, and clinically meaningful, rather than being driven by a small number of favorable scenarios. It further supports the claim that structured state-aware multi-agent collaboration is particularly effective for complex, multi-system clinical intake tasks.

\subsection{Case Study: High-Risk Prostate Cancer}
\label{subsec:case_study}
\begin{figure*}
    \centering
    \includegraphics[width=1\linewidth]{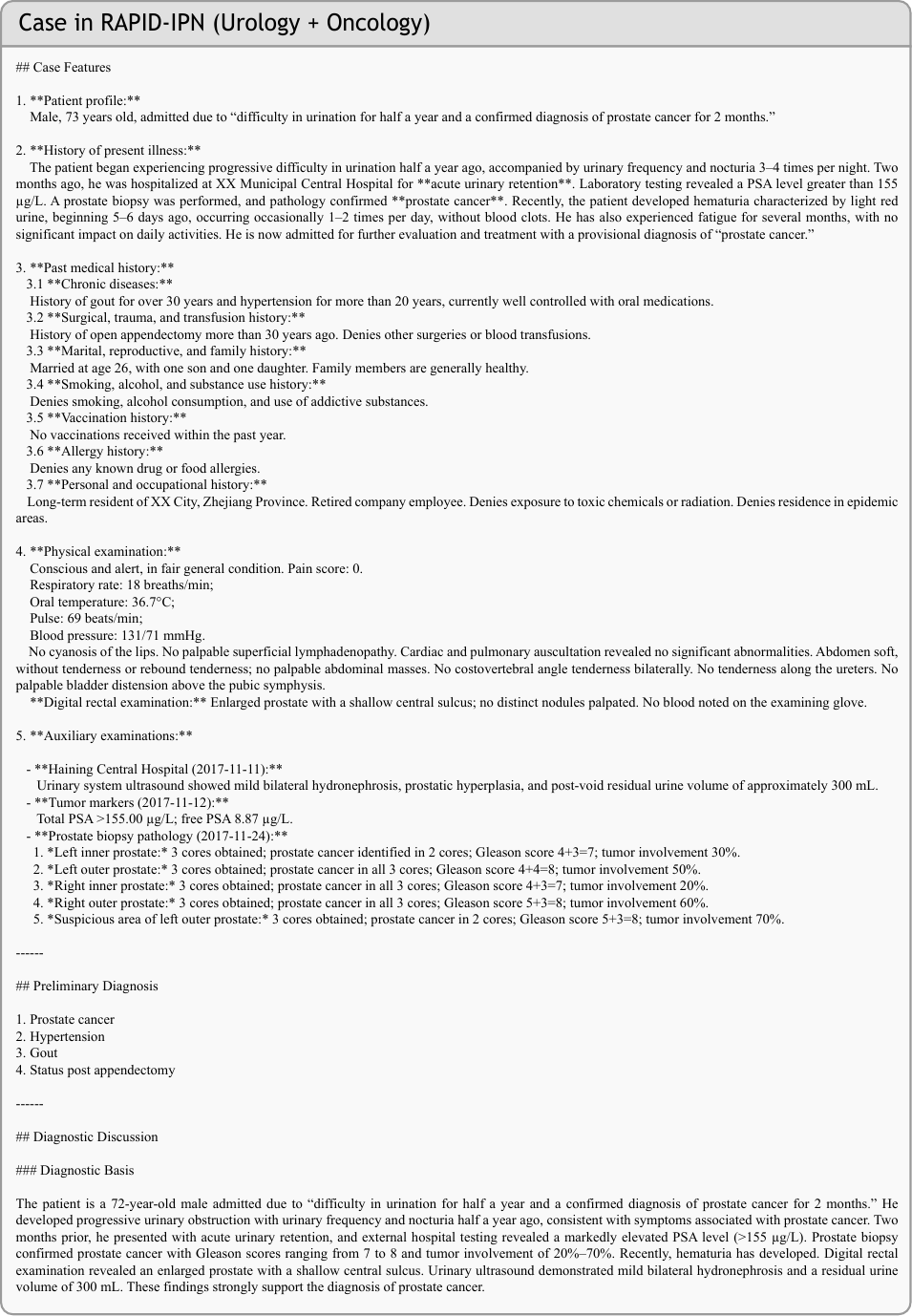}
    \caption{The IPN of a 73-year-old male patient suffering from prostate cancer.}
    \label{fig:case_study_1}
\end{figure*}

\begin{figure*}
    \centering
    \includegraphics[width=1\textwidth]{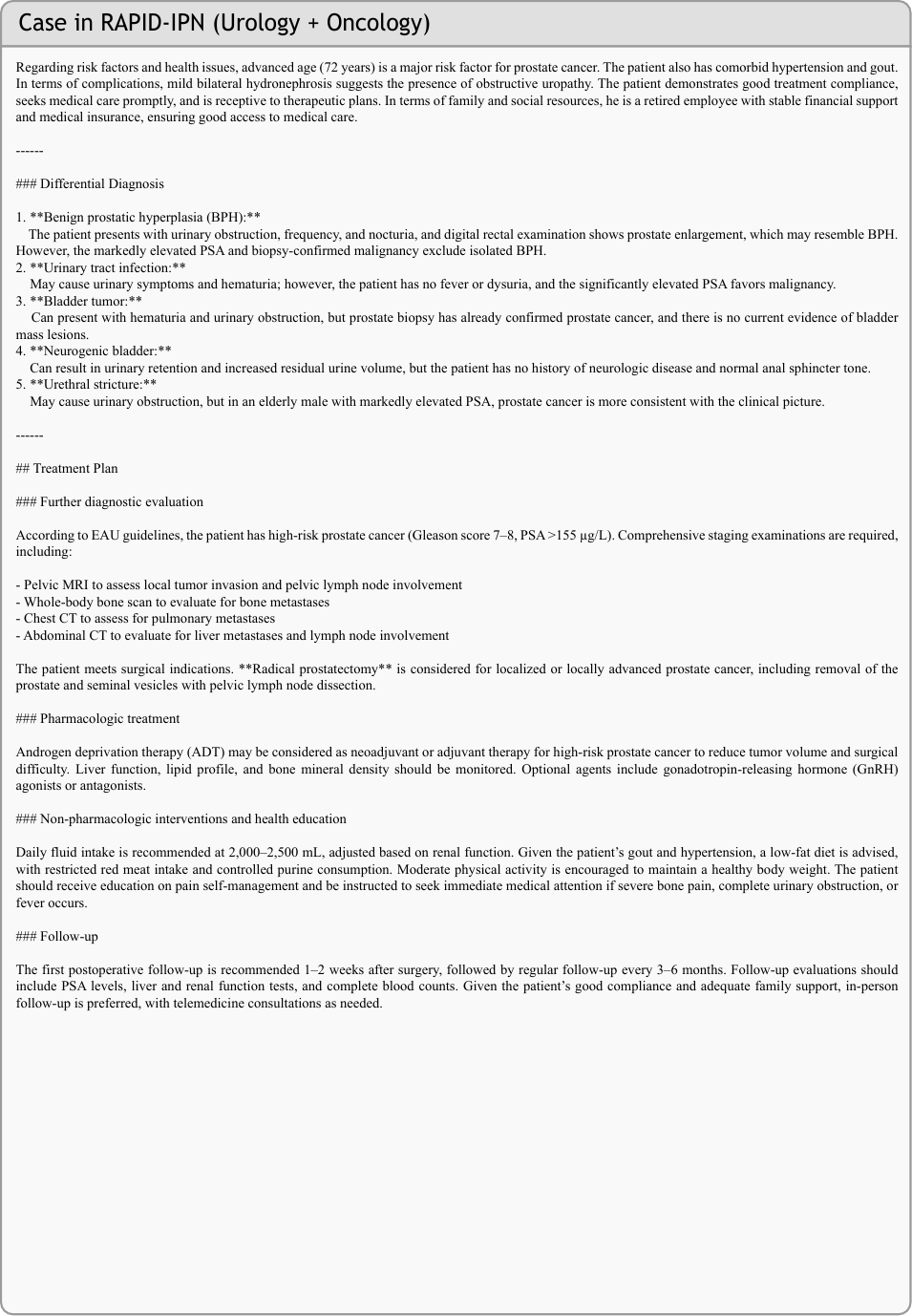}
    \caption{The IPN of a 73-year-old male patient suffering from prostate cancer (continued).}
    \label{fig:case_study_2}
\end{figure*}

To qualitatively demonstrate the advantages of Aegle’s virtual MDT framework, we analyze a complex real-world case from the RAPID-IPN dataset involving a 73-year-old male presenting with progressive lower urinary tract symptoms (LUTS) and a PSA level $>155$ ng/mL (Fig~\ref{fig:case_study_1} and Fig.~\ref{fig:case_study_2}). This case requires integrating urological history, oncology pathology, and imaging evidence to formulate a high-risk management plan.

\paragraph{Precision in Evidence Acquisition.}
As illustrated in Table~\ref{tab:key_clinical_evidence}, the primary challenge in this case was not the diagnosis of prostate cancer, which had already been confirmed by biopsy, but the accurate characterization of risk stratification and the severity of urinary obstruction. Reasoning-strategy baselines (CoT and ToT) captured high-level symptoms but failed to record granular metrics required for surgical planning. Specifically, both CoT and ToT omitted the quantitative International Prostate Symptom Score (IPSS) and the exact blood pressure measurement documented during the physical examination. In contrast, Aegle’s dynamic topology activated a specialized Urologist Agent during the inquiry phase (Stage I). Whereas MDAgents failed to capture the critical urinary retention metric that determines the urgency of decompression, Aegle successfully incorporated this information into the \textit{Objective} section of the IPN.
\begin{table*}[ht]
\centering
\resizebox{\textwidth}{!}{
\begin{tabular}{lccccc}
\toprule
\textbf{Key Clinical Evidence} 
& \textbf{Aegle} 
& \textbf{CoT} 
& \textbf{ToT} 
& \textbf{MDAgents} 
& \textbf{MedAgents} \\
\midrule

Age: 73-year-old male 
& $\checkmark$ & $\checkmark$ & $\checkmark$ & $\checkmark$ & $\checkmark$ \\

Progressive LUTS (weak stream, hesitancy, intermittency, straining, dribbling) 
& $\checkmark$ & $\checkmark$ & $\checkmark$ & $\checkmark$ & $\checkmark$ \\

Nocturia 3--4 times/night 
& $\checkmark$ & $\checkmark$ & $\checkmark$ & $\checkmark$ & $\checkmark$ \\

Intermittent painless light-red hematuria (1--2$\times$/day, no clots) 
& $\checkmark$ & $\checkmark$ & $\checkmark$ & $\checkmark$ & $\checkmark$ \\

Acute urinary retention 2 months ago (catheterized) 
& $\checkmark$ & $\checkmark$ & $\checkmark$ & $\checkmark$ & $\checkmark$ \\

PSA $>$155 ng/mL 
& $\checkmark$ & $\checkmark$ & $\checkmark$ & $\checkmark$ & $\checkmark$ \\

Prostate biopsy: adenocarcinoma, Gleason 7--8, multifocal, 20--70\% involvement 
& $\checkmark$ & $\triangle$ & $\triangle$ & $\triangle$ & $\checkmark$ \\

Urinary ultrasound: bilateral mild hydronephrosis 
& $\checkmark$ & $\checkmark$ & $\checkmark$ & $\checkmark$ & $\checkmark$ \\

Post-void residual $\approx$300 mL 
& $\checkmark$ & $\checkmark$ & $\checkmark$ & $\times$ & $\checkmark$ \\

No CT / MRI / bone scan performed yet 
& $\checkmark$ & $\checkmark$ & $\checkmark$ & $\triangle$ & $\checkmark$ \\

Renal function \& baseline labs not yet available 
& $\checkmark$ & $\checkmark$ & $\triangle$ & $\checkmark$ & $\checkmark$ \\

Blood pressure from original exam
& $\checkmark$ & $\times$ & $\checkmark$ & $\checkmark$ & $\triangle$ \\

IPSS score mentioned 
& $\checkmark$ & $\times$ & $\times$ & $\times$ & $\times$ \\

\bottomrule
\end{tabular}
}
\caption{Coverage of key clinical evidence across different reasoning frameworks.
$\checkmark$ = explicitly documented; 
$\triangle$ = partially mentioned or ambiguous; 
$\times$ = missing;
}
\label{tab:key_clinical_evidence}
\end{table*}

\paragraph{Handling Diagnostic Ambiguity.}
The pathology report described multifocal adenocarcinoma with heterogeneous Gleason scores across biopsy cores. Standard baselines generally summarized these findings as ``Prostate Cancer'' without further differentiation. MedAgents and Aegle were the only models that retained core-level involvement percentages ranging from 20 to 70\%. Importantly, Aegle extended beyond data retention in the \textit{Assessment} section by synthesizing these findings to correctly classify the patient as high risk according to EAU guidelines. This clinically meaningful distinction was not captured by the generic summarization produced by CoT.

\newpage
\paragraph{Plan Coherence.}
The improved evidence grounding directly translated into a higher-quality treatment plan. Because Aegle explicitly encoded the markedly elevated PSA level ($>155$) and the Gleason score of 8 within the clinical state $\mathcal{F}$, the \textit{Oncologist Agent} in Stage II generated a comprehensive staging strategy. This plan included a whole-body bone scan and pelvic MRI to exclude metastatic disease prior to scheduling radical prostatectomy. This case demonstrates how Aegle’s decoupled reasoning framework preserves low-salience yet high-impact clinical details, ensuring that the final IPN satisfies the standards required for specialist referral.

\end{document}